\documentclass[%
reprint,
superscriptaddress,
preprintnumbers,
amsmath,
amssymb,
aps,
floatfix,
]{revtex4-2}
\usepackage{graphicx}
\usepackage{dcolumn}
\usepackage{bm}
\usepackage[utf8]{inputenc}
\usepackage[T1]{fontenc}
\usepackage{mathptmx}
\usepackage{textcomp}
\usepackage{siunitx}
\usepackage{bigints}

\usepackage[switch]{lineno}
\usepackage[table,xcdraw,dvipsnames]{xcolor}
\usepackage{booktabs}
\usepackage{float}
\usepackage{color}
\usepackage{ulem}
\usepackage{multirow}

\begin{document}
\title[Skyrmionic device for three dimensional magnetic field sensing enabled by spin-orbit torques]{Skyrmionic device for three dimensional magnetic field sensing enabled by spin-orbit torques}

\author{Sabri~Koraltan}
\email{sabri.koraltan@univie.ac.at}
\affiliation{Physics of Functional Materials, Faculty of Physics, University of Vienna, A-1090 Vienna, Austria}%
\affiliation{Vienna Doctoral School in Physics, University of Vienna, Vienna, Austria}%
\affiliation{Research Platform MMM Mathematics-Magnetism-Materials, University of Vienna, A-1090 Vienna, Austria.}%

\author{Rahul~Gupta}
\affiliation{Institute of Physics, Johannes Gutenberg University Mainz, 55099 Mainz, Germany}%

\author{Reshma~Peremadathil~Pradeep}
\affiliation{Empa, Swiss Federal Laboratories for Materials Science and Technology, Dübendorf CH-8600, Switzerland}

\author{Fabian~Kammerbauer}
\affiliation{Institute of Physics, Johannes Gutenberg University Mainz, 55099 Mainz, Germany}%

\author{Iryna Kononenko}
\affiliation{Institute of Physics, Johannes Gutenberg University Mainz, 55099 Mainz, Germany}%
\affiliation{National Academy of Sciences of Ukraine, Institute of Applied Physics, 58 Petropavlivska St., Sumy, 40000, Ukraine}

\author{Klemens~Prügl}%
\affiliation{Infineon Technologies AG, 93049 Regensburg, Germany}

\author{Michael~Kirsch}%
\affiliation{Infineon Technologies AG, 93049 Regensburg, Germany}

\author{Bernd~Aichner}
\affiliation{Faculty of Physics, University of Vienna, A-1090 Vienna, Austria}%

\author{Santiago~Helbig}
\affiliation{Faculty of Physics, University of Vienna, A-1090 Vienna, Austria}%

\author{Florian~Bruckner}
\affiliation{Physics of Functional Materials, Faculty of Physics, University of Vienna, A-1090 Vienna, Austria}%

\author{Claas~Abert}%
\affiliation{Physics of Functional Materials, Faculty of Physics, University of Vienna, A-1090 Vienna, Austria}%
\affiliation{Research Platform MMM Mathematics-Magnetism-Materials, University of Vienna, A-1090 Vienna, Austria.}%

\author{Andrada~Oana~Mandru}
\affiliation{Empa, Swiss Federal Laboratories for Materials Science and Technology, Dübendorf CH-8600, Switzerland}

\author{Armin~Satz}
\affiliation{Infineon Technologies AG, 9500 Villach, Austria}

\author{Gerhard~Jakob}
\affiliation{Institute of Physics, Johannes Gutenberg University Mainz, 55099 Mainz, Germany}%

\author{Hans~Josef~Hug}
\affiliation{Empa, Swiss Federal Laboratories for Materials Science and Technology, Dübendorf CH-8600, Switzerland}
\affiliation{Department of Physics, University of Basel, 4056 Basel, Switzerland}

\author{Mathias~Kl\"{a}ui}
\affiliation{Institute of Physics, Johannes Gutenberg University Mainz, 55099 Mainz, Germany}%

\author{Dieter~Suess}%
\affiliation{Physics of Functional Materials, Faculty of Physics, University of Vienna, A-1090 Vienna, Austria}%
\affiliation{Research Platform MMM Mathematics-Magnetism-Materials, University of Vienna, A-1090 Vienna, Austria.}%

\date{\today}

\begin{abstract}
Magnetic skyrmions are topologically protected local magnetic solitons that are promising for storage, logic or general computing applications. In this work, we demonstrate that we can use a skyrmion device based on $\mathrm{[W/CoFeB/MgO]}_{10}$ multilayers for three-dimensional magnetic field sensing enabled by spin-orbit torques (SOT). We stabilize isolated chiral skyrmions and stripe domains in the multilayers, as shown by magnetic force microscopy images and micromagnetic simulations. We perform magnetic transport measurements to show that we can sense both in-plane and out-of-plane magnetic fields by means of a differential measurement scheme in which the symmetry of the SOT leads to cancelation of the DC offset. With the magnetic parameters obtained by vibrating sample magnetometry and ferromagnetic resonance measurements, we perform finite-temperature micromagnetic simulations, where we investigate the fundamental origin of the sensing signal. We identify the topological transformation between skyrmions, stripes and type-II bubbles that leads to a change in the resistance that is read-out by the anomalous Hall effect. Our study presents a novel application for skyrmions, where a differential measurement sensing concept is applied to quantify external magnetic fields paving the way towards more energy efficient applications in skyrmionics based spintronics.

\end{abstract}
\maketitle

\subsection*{\label{sec:level0}Introduction}
Magnetic skyrmions are localized magnetic solitons that are topologically protected~\cite{roessler2006spontaneous, muehlbauer2009skyrmion, finocchio2016magnetic}. Their topological protection means that they cannot be annihilated into the ferromagnetic state by a purely uniform and continuous transformation without the appearance of magnetic singularites~\cite{buttner2018theory}. Skyrmions can have different stabilization mechanisms. Most commonly, they are found in systems where the inversion symmetry is broken due to the presence of the Dzyaloshinskii-Moriya interaction(DMI) . Multiple systems exist where bulk~\cite{yu2010real} or interfacial DMI~\cite{romming2013writing} is responsible for the formation of skyrmions. The latter relies on the DMI that arises at the interface between a ferromagnetic layer and heavy metal layers~\cite{kuepferling2023measuring}. Thus, metallic multilayers can be used to enhance the stability region of skyrmions, allowing them to also appear at room temperature~\cite{soumyanarayanan2017tunable}. These chiral skyrmions are called Néel skyrmions as their cross-section reveals a Néel domain wall~\cite{buttner2018theory}. Note that skyrmions can also be stabilized in materials without DMI, where the competition between demagnetizing energies and low perpendicular magnetic anisotropy leads to the formation of Bloch skyrmions\cite{montoya2017tailoring, heigl2021dipolar} and even higher-order skyrmions and antiskyrmions~\cite{hassan2024dipolar}. 

Because the skyrmions can be found over a wide range of temperatures where their sizes depend on the applied external fields and because of their energy efficient propagation by electric currents~\cite{fert2017magnetic, woo2016observation}, they have been proposed as information carriers in skyrmion race track memories~\cite{fert2013skyrmions, tomasello2014strategy}, logic devices~\cite{luo2018reconfigurable}, spin-wave emitters~\cite{chen2021chiral, chumak2022advances} and even in unconventional computing tasks~\cite{song2020skyrmion,finocchio2023roadmap} such as reservoir computers~\cite{prychynenko2018magnetic, lee2024task, raab2022brownian}. The latter is based on the change in resistance that is obtained when skyrmions are breathing or propagating. Generally, a transverse voltage results from the electrons flowing in a skyrmionic circuit~\cite{guang2023electrical}. The most common contributions are the anomalous Hall effect (AHE)~\cite{nagaosa2010anomalous} and topological Hall effect (THE)~\cite{neubauer2009topological}. While the former arises due to the change in the magnetization (for instance, breathing of skyrmions), the latter emerges due to the presence of skyrmions as topological defects that deflect the electrons. Thus, AHE and THE can be used to characterize the change in the magnetization of a skyrmionic system.

Recently, AHE-based magnetic field sensing devices have triggered increased attention~\cite{zhu2014giant,peng2019ultrasensitive,koraltan2023single}. 
The concept of magnetic field sensing can be summarized quite simply: The aim is to measure the change in resistance when the externally applied magnetic field changes~\cite{lenz1990review}. Most designs rely on Hall effect or magnetoresistive effects: anisotropic magnetoresistance (AMR), giant magnetoresistance (GMR), or tunnel magnetoresistance (TMR)~\cite{freitas2007magnetoresistive}. TMR sensors are indeed gaining popularity in various applications due to their high sensitivity compared to traditional Hall effect sensors. The drawback of state-of-the-art TMR sensors is that they rely only on the shape anistropy of an ellipsoidal element, where only small linear ranges can be achieved, strong hysteretic effects appear, while high phase noise can be measured. Novel concepts have been proposed and even commercially adapted, where magnetic vortices were used to overcome these drawbacks~\cite{suess2018topologically}.
Recently, AHE-based sensors have been shown to be highly sensitive~\cite{zhu2014giant,peng2019ultrasensitive}, still lower than the sensitivities achieved by TMR sensors. In most AHE-based sensors, the measurable linear ranges do not exceed $\SI{4}{mT}$, which limits the application range of such sensors~\cite{zheng2019magnetoresistive}.

Further research work focussed on the use of spin-orbit torques ~\cite{mihai2010current, miron2011perpendicular, gambardella2011current} (SOT) for an energy-efficient switching of the magnetization in prototypical devices. Additional work used SOT for the controlled propagation of a domain wall for three-dimensional field sensing~\cite{li2021spin}, whereas our own work demonstrated the suppression of inevitable sensor offsets by SOT~\cite{koraltan2023single}. However, while the linear field range remained limited for the 3D field sensing devices~\cite{li2021spin}, the sensor concepts developed in our own work suffered from a relatively small field sensitivity~\cite{koraltan2023single}.

In this work, we design and fabricate a skyrmionic hall bar type device to measure in-plane (IP) and out-of-plane (OOP) components of the applied magnetic field. The electrical read out makes use of the anomalous Hall effect. Magnetic force microscopy performed under vacuum conditions is employed to reveal the evolution of stripe domains to skyrmions and to test the fidelity of micromagnetic simulations. We then use the latter to demonstrate that the formation or annihilation of skyrmions, trivial bubbles and stripe domains becomes symmetric with respect to the applied current direction due to the SOT, which enables offset cancelation for IP fields. Micromagnetic simulations accompany our experimental studies and reveal the origin of the offset-free sensing principle and how the SOT introduced a symmetry in the orientation of stripe domains and formation of spin textures. We achieve a measurable linear range of $\SI{\pm 17}{mT}$ for IP fields and $\SI{\pm 30}{mT}$ for OOP fields, respectively, while keeping the sensitivity moderately high. Our sensing performance exceeds both previous academic and commercial devices, opening a new paradigm for research in skyrmions-based spintronics.

\subsection*{\label{sec:level1}Néel Skyrmions in W/CoFeB/MgO Multilayers}
\begin{figure*}
    \centering
    \includegraphics[width=\textwidth]{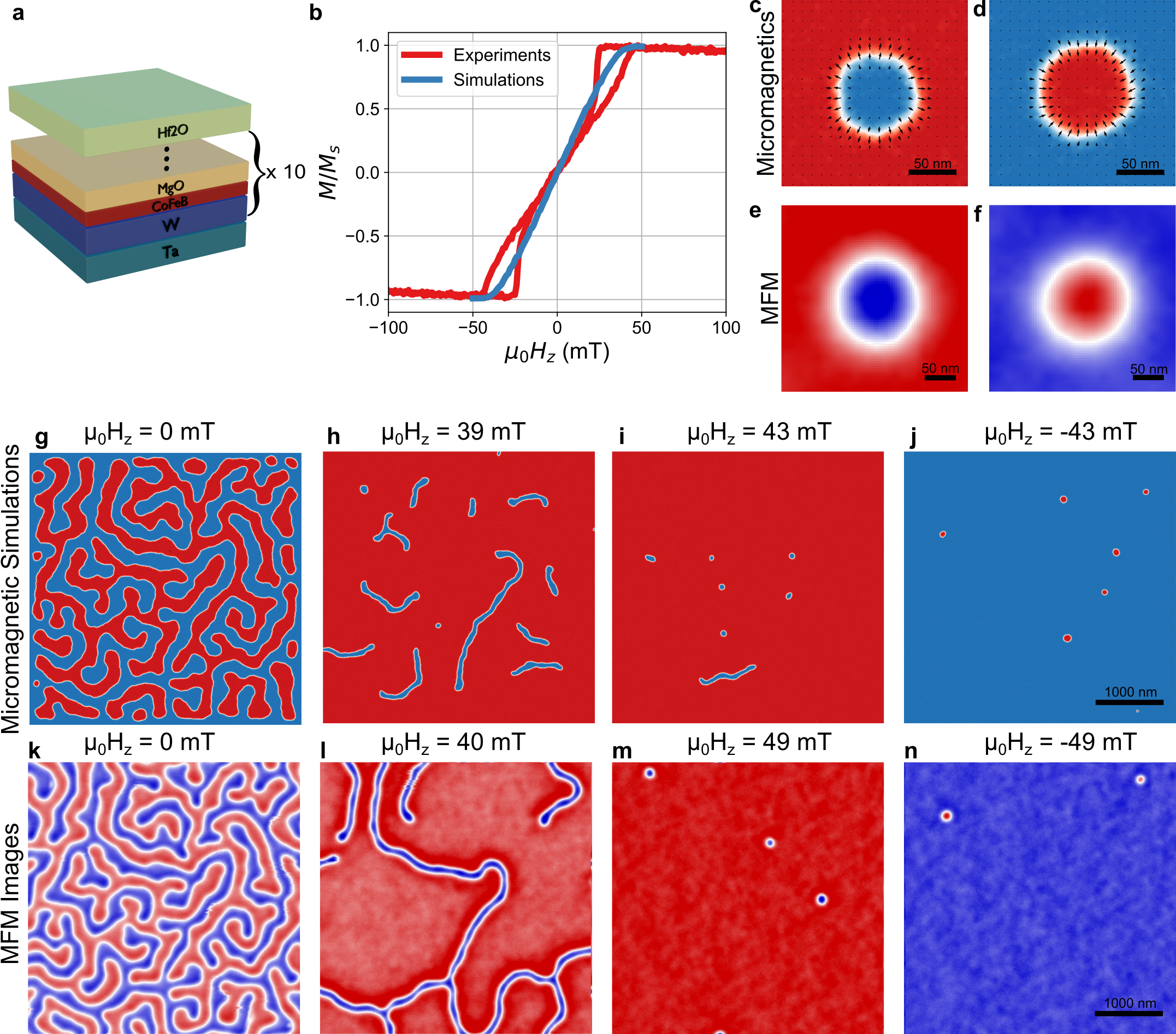}
    \caption{\textbf{Formation of Néel Skyrmions.} The investigated thin-film [W(5)/CoFeB(0.7)/MgO(1.2)]$_{\mathrm{N}}$ multilayer stack is illustrated in (a) where the layers are grown on a Ta(5) seed layer and capped with $\mathrm{HfO_2}$ (3) layer to avoid oxidation. The numbers in parantheses represent the thickness of each layer in nanometers. The hysteresis curve experimentally acquired by VSM (red) and micromagnetically simulated curve (blue) are compared in (b). Isolated Néel skyrmions as obtained from micromagnetic simulations for negative (c) and positive (d) skyrmion cores, and as obtained from MFM images in (e) and (f), respectively. The evolution of the magnetization states for different magnetic fields is illustrated in (g)-(j) as snapshots of micromagnetic simulations and as experimentally imaged using MFM, showing the measured frequency shift patterns arising from the magnetization patterns in (k)-(n). From the MFM images at vanishing fields, an average domain width $\sigma_{DW} = \SI{166.34}{nm} \pm \SI{22.56}{nm}$ was calculated. The micromagnetic simulations use the input of the MFM state as an initial state before applying the OOP field to observe the formation of skyrmions.}
    \label{fig:fig01}
\end{figure*}
The interfacial DMI originating from the spin-orbit coupling arising at the interface between a heavy metal and ferromagnetic layers is a well-studied mechanism for the nucleation of skyrmions under an external field~\cite{soumyanarayanan2017tunable, jaiswal2017investigation}. In this article, we study a multilayer Si/SiO${_2}$/Ta(5)/[W(5)/CoFeB(0.7)/MgO(1.2)]x10 sample (all thicknesses in nanometers), which was deposited by DC and RF magnetron sputtering (as depicted in Fig.~\ref{fig:fig01}a); see Methods for more details. 

In our study, the relevant system parameters of our magnetic thin films were analyzed through the application of vibrating sample magnetometry (VSM), ferromagnetic resonance (FMR), and magnetic force microscopy (MFM). 
A saturation magnetization $M_s = \SI{1197.5}{kA/m}$ was determined by VSM and then used to determine the anisotropy constant $K_u = \SI{1042}{kJ/m^3}$ from frequency of the Kittel resonance mode~\cite{kittel1948theory, beaujour2007ferromagnetic} obtained from FMR.
For the FMR measurements, the samples were capped by a Hf2O to mitigate electromagnetic interference with the coplanar waveguide. Detailed methodologies of these measurements are elaborated in the Methods section. 
Note that the anisotropy originates from the W/CoFeB/MgO interfaces~\cite{ikeda2010perpendicular, cui2013perpendicular}. The obtained anisotropy and saturation magnetization agree well with previously reported values~\cite{jaiswal2017investigation, bottcher2021heisenberg} for similar systems. 
We successfully derived temperature-dependent material parameters via VSM and FMR, which we subsequently used for our micromagnetic simulations conducted on \texttt{magnum.np}~\cite{bruckner2023magnumnp}. These measurements are depicted in Extended Data Figure~\ref{fig:EDFig01} and Extended Data Figure~\ref{fig:EDFig02} for VSM and FMR outcomes, respectively. For futher simulations, the exchange constant $A_\mathrm{ex}$ was taken from literature~\cite{bottcher2021heisenberg}. The value for DMI was numerically optimized to obtain the same zero field domain morhopology as measured by MFM; see Methods and Extended Data Fig.~\ref{fig:EDFig03} for more details. 

In Fig.~\ref{fig:fig01}b we depict the response of magnetization to an out-of-plane (OOP) field (MH-Loops) as measured with VSM (red) and as simulated (blue) at room temperature. We observe the two hysteresis pockets, where the system collapses into a multi-domain/striped domain phase from which skyrmions can form. To directly observe the existence of skyrmions in our samples, we acquired magnetic force microscopy (MFM)~\cite{hug1998quantitative, feng2022quantitative} images for different magnetic fields applied in the OOP direction. Additionally, we accompany our experimental findings with snapshots of the magnetization states from micromagnetic simulations, where we used the zero-field MFM state as the initial state. Snapshots of the magnetization for the simulation results of the blue curve in Fig.~\ref{fig:fig01}b are given in Extended Data Fig.~\ref{fig:EDFig03}, where a high density of skyrmions is observed if one starts from a random magnetization state for each field. However, this represents rather a local energy minimum, and in reality, a sparse skyrmion nucleation is obtained. Isolated Néel skyrmions are depicted for both core polarities from simulations (Fig.~\ref{fig:fig01}c,d) and MFM experiments (Fig.~\ref{fig:fig01}e,f). The micromagnetic simulations (Fig.~\ref{fig:fig01}g-j) and MFM images (Fig.~\ref{fig:fig01}k-n) reveal that a rich multidomain (stripe) state is achieved at zero field. By performing a domain width analysis, we find the average domain width $\sigma_{DW} = \SI{166.34}{nm} \pm \SI{22.56}{nm}$. Note that we used the magnetization state obtained from MFM as the initial state in our micromagnetic simulations, which allows for an exceptionally good agreement. If the magnetic field is increased, the stripes start to shrink, and eventually Néel skyrmions form for sufficiently large fields ($> \SI{40}{mT}$). In general, we confirmed the presence of skyrmions in our sample by employing MFM imaging and micromagnetic simulations, where we found skyrmions with an average radius in the range of $r_{Sk} \approx \SI{60}{nm}$. Further MFM images at different magnetic fields are provided in the Extended Data Fig.~\ref{fig:EDFig04}

.

\subsection*{\label{sec:level2}Three Dimensional Magnetic Field Sensor}
\begin{figure*}
    \centering
    \includegraphics[width=\textwidth]{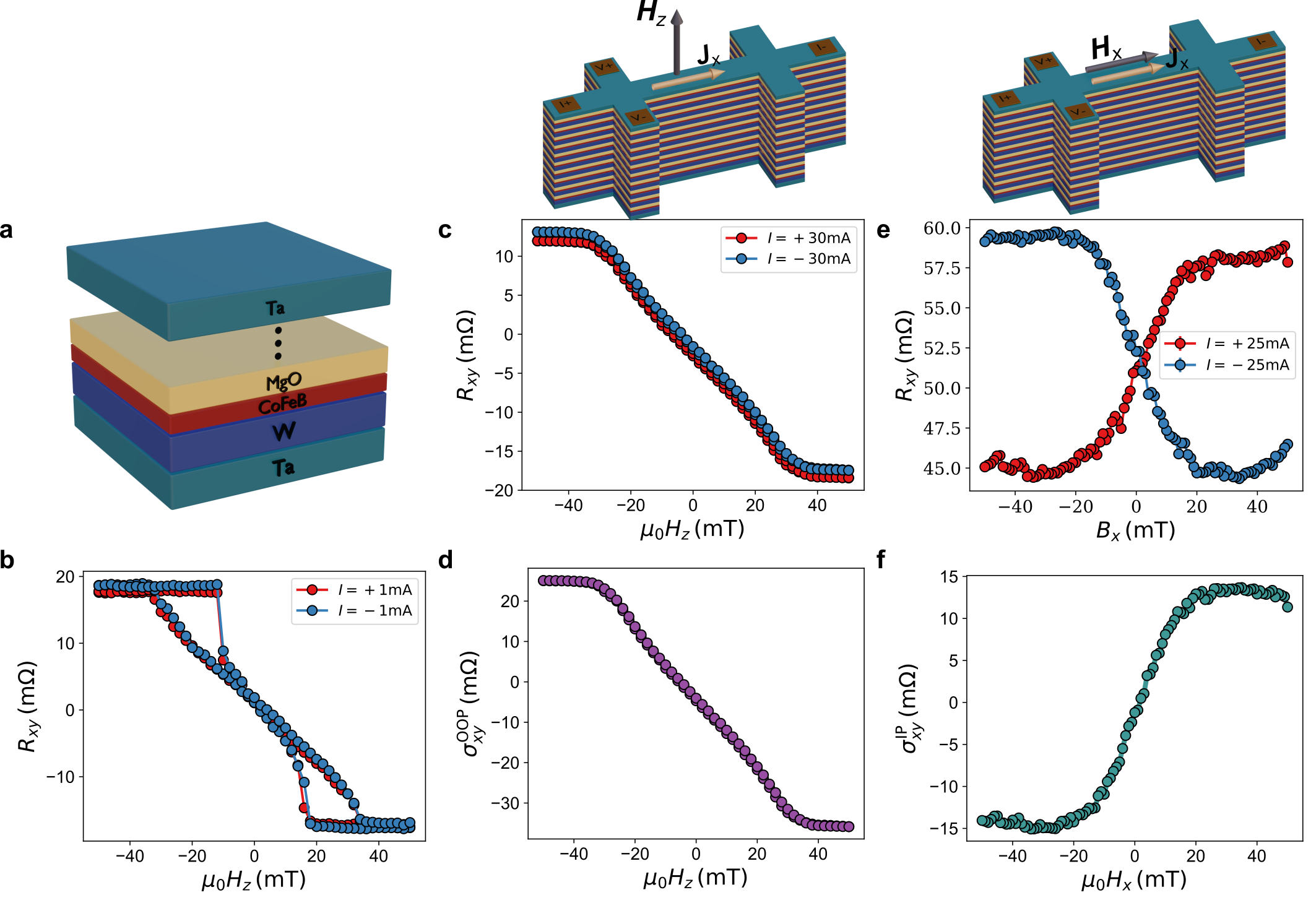}
    \caption{\textbf{Three dimensional sensing using a skyrmionic device.} The multilayer stack (a) used to pattern the Hall bars shown artistically in the upper pannels, where the contacting for a positive flowing current and the direction of the recorded transverse voltage are highlighted. In contrast to thin films, we use $\mathrm{Ta}$(3) as a capping layer. The transverse resistance $R_{xy}$ as a function of the applied OOP magnetic field for both current polarities $I=\SI{\pm 1}{mA}$ (b) reproducing the VSM curve from Fig.~\ref{fig:fig01}b. The change in $R_{xy}$ when an OOP field is applied under a strong injection of SOT current ($I=\SI{30}{mA}$) in (c), where the hysteresis from (b) vanishes and a very linear signal $\sigma^\mathrm{OOP}_{xy} = R_{xy}(+I) + R_{xy}(-I)$ can be obtained by summing the two curves for the two current polarities (d), demonstrating that the skyrmionic device can be used to sense OOP magnetic fields. The evolution of $R_{xy}$ when an IP field is applied for positive and negative currents of magnitude $|I| = \SI{25}{mA}$ is depicted in (e), where the symmetry of the SOT reverses the curves, allowing us to employ the offset-free sensing principle to obtain the sensor transfer curve (f) where $\sigma^\mathrm{IP}_{xy} = R_{xy}(+I) - R_{xy}(-I)$ is quantified. All measurements are performed at an environmental temperature $T=\SI{300}{K}$.}
    \label{fig:fig02}
\end{figure*}

In the following, we discuss how our skyrmionic device can be used as a magnetic-field sensor to sense all three directions. For this purpose, we use Si/SiO{$_2$}/Ta(5)/[W(5)/CoFeB(0.7)/MgO(1.2)] $\times$ 10/Ta(3) multilayers that have also been grown by DC magnetron sputtering; see Fig.~\ref{fig:fig02}a. We structured 6-armed Hall bars using convential photolithography and $\mathrm{Ar^{+}}$ ion etching using a hard mask. More details are given in the Methods section. The Hall bar has a width of $w=\SI{2}{\mu m}$. Throughout this paper, we follow a differential DC measurement protocol. That is, the current is first injected along the positive direction and the longitudinal ($R_{xx}$) and transverse ($R_{xy}$) resistances are recorded. Then, the current polarity is reversed and the resistances are recorded again. Only after the resistances for both current directions have been recorded, the externally applied magnetic field is changed.

We start the transport measurements by injecting a low-amplitude current $I = \SI{1}{mA}$ into both positive and negative $x$ directions and recording the transverse resistance ($R_{xy}$) as a function of OOP field, as depicted in Fig.~\ref{fig:fig02}b. More details of the measurements are given in the Methods section. When the injected currents are small, the current density is insufficient to induce the motion of the skyrmions or change the magnetization state. Furthermore, the impact of SOT is negligible. This allows us to qualitatively analyze conventional MH loops electrically. The experimental data shown in Fig.~\ref{fig:fig02}b closely match the MH loops obtained previously by VSM. It is important to note that both AHE and THE could potentially contribute to $R_{xy}$. Although THE could be quantified from this signal, it is beyond the scope of this study. The electrical signal obtained can already function as a sensor signal, as changes in resistance are observed as a result of the shrinking of stripes and subsequent formation of skyrmions. The linear measurement range of the sensor is $LR = \SI{\pm 10}{mT}$. A slight hysteresis is noticeable, possibly indicating variations in the population of the multi-domain state with stripes when returning from a saturated state.

The overall scenario undergoes a substantial transformation when the amplitude of the injected current is greatly increased. As the current density approaches orders of magnitude close to electromigration, the systems temperature increases significantly due to Joule heating. To understand how the system heats up with respect to the injected in-plane current, we perform a series of transport measurements in which (I) we quantify the longitudinal resistance $R_{xx}$ as a function of environmental temperature, and (II) we measure $R_{xx}$ for different injected currents at room temperature while the system was magnetically saturated. We then correlate the two measurements to obtain the sample temperature as a function of the applied current; see Extended Data Fig.~\ref{fig:EDFig05}. For $I = \SI{\pm 30}{mA}$ we deduced the sample temperature $T_{\mathrm{sample}} = \SI{562.9}{K}$. For the calculation of the current density, we assume that the current flow is homogeneous in all layers. Taking into account the cross section of current flow with thickness $t = \SI{74}{nm}$ and width $w = \SI{2}{\mu m}$, the resulting current density is $j_e \approx \SI{2e11}{A/m^2}$. This current density is sufficient to harness very strong effective SOT fields, which can be summarized as field-like ($\mathbf{T}_\mathrm{fl}$) and damping-like ($\mathbf{T}_\mathrm{dl}$) torques that extend the regular LLG equation~\cite{manchon2019current,abert2019micromagnetics}. Both the damping-like torque (DLT) and the field-like torque (FLT) depend on the local direction of the spin polarization $\mathbf{p}$ and of the magnetization $\mathbf{m}$, and can be expressed via

\begin{equation}
    \mathbf{T}_\mathrm{DL} = \mathbf{m}\times\mathbf{H}_{\mathrm{DL}} = \eta_\mathrm{DL}\dfrac{j_e\gamma\hbar}{2e\mu_0tM_s}{\mathbf{m}\times\left( \mathbf{m} \times \mathbf{p}\right)},
    \label{eq:eq01}
\end{equation}

and
\begin{equation}
    \mathbf{T}_\mathrm{FL} = \mathbf{m}\times\mathbf{H}_{\mathrm{FL}} = \eta_\mathrm{FL}\dfrac{j_e\gamma\hbar}{2e\mu_0tM_s}{\mathbf{m} \times \mathbf{p}},
    \label{eq:eq02}
\end{equation}

where $\gamma = \SI{2.21e5}{(m/As)}$ represents the gyromagnetic ratio, $\hbar$ denotes the reduced Planck constant, $e$ stands for the elementary charge, and $t$ signifies the thickness of the ferromagnetic layer. The effective SOT fields $\mathbf{H}_\mathrm{DL}$ and $\mathbf{H}_\mathrm{FL}$ directly enter the LLG effective field term, where their magnitude is given by the dimensionless SOT coefficients $\eta_\mathrm{DL}$ and $\eta_\mathrm{FL}$, respectively. The electrons injected into the system acquire spin polarization as a result of the spin Hall~\cite{sinova2015spin} and Rashba-Edelstein~\cite{mihai2010current} effects. When the current is injected along the $x$ axis, and the spin-polarized current flows into the skyrmionic device following the $+z$ direction, the electrons acquire spin polarization along $-y$~\cite{manchon2019current, koraltan2023single}.

Now, we increase the applied current to $I = \SI{\pm 30}{mA}$. The experimentally recorded $R_{xy}$ as a function of the applied OOP field $\mu_0H_z$ is depicted in Fig.~\ref{fig:fig02}c for both positive and negative current directions. Compared to the VSM hysteresis loops, or low-current measurements, we observe that the hysteresis and the characteristic pockets disappear, allowing us to harness a highly linear and hysteresis-free magnetic signal. By adding the resistances for positive and negative currents, a highly linear sensing signal $\sigma^\mathrm{OOP}_xy(H_z) = R_{xy}(+I) + R_{xy}(-I)$ is obtained, as shown in Fig.~\ref{fig:fig02}d. The measurable linear range increases to $LR \approx \pm \SI{30}{mT}$, while the sensitivity is quantified as $S_z = \dfrac{\mathrm{d}^2V_{xy}}{\mu_0\mathrm{d}H_z\mathrm{d}I} = \dfrac{\mathrm{d}\sigma^{\mathrm{OOP}}_{xy}}{\mu_0\mathrm{d}H_z} \approx \SI{0.8}{(V/A)/T}$. Note that we do not average the two curves, thus, the magnitude of the sensing signal is twice as high as individual transfer curves. In Extended Data Fig.~\ref{fig:EDFig06} we demonstrate how the hysteresis decreases with increasing injected current and thus leads to an improvement in the measurable linear range of the sensing signal. As for many applications, it is important that the sensors also work at lower temperatures; we repeated our experiments at lower temperatures ($T=\SI{200}{K}$; see again Extended Data Fig.~\ref{fig:EDFig06}). When the environmental temperature is decreased, the magnetic parameters, e.g. $M_s$, and $K_u$, increase, while the effective SOT fields induced by current are not significantly affected, as the change in the magnetic parameters can be compensated by the temperature dependence of the SOT coefficients. Overall, our measurements indicate that our skyrmionic device can be utilized to sense OOP magnetic fields with a high linear working range and moderate sensitivities.

So far we have considered only OOP fields, while the current was applied along the $x$-axis. We now change our focus to in-plane (IP) magnetic fields, which are always applied parallel to the injected current. In Fig.~\ref{fig:fig02}e we compare the measured resistances for a positive current (red) and a negative current (blue). The two curves cross each other at vanishing external fields. When comparing these curves to those under OOP fields, we understand following behavior: while a positive current is injected, a positive (negative) IP field strives for achieving negative (positive) magnetic saturation. By reversing the current polarity, now a positive fields tries to reach positive saturation. For sufficiently high IP magnetic fields, it is possible to saturate the magnetization OOP, before turning the entire system IP, when $H\approx H_k$. As in our previous work~\cite{koraltan2023single}, we exploit this symmetry and our differential measurement protocol allows us to obtain a sensing signal ($\sigma^\mathrm{IP}_{xy}(H_x) = R_{xy}(+I) - R_{xy}(-I)$) for IP magnetic fields. Due to the symmetry of the SOT, the sensing offset is nearly eliminated. Hence, our skyrmionic device can also be used to sense IP magnetic fields with a linear range of $LR \approx  \pm \SI{17}{mT}$, negligible DC offset, and a sensitivity ($ S_x = \dfrac{\mathrm{d}^2V_{xy}}{\mu_0\mathrm{d}H_x\mathrm{d}I} = \dfrac{\mathrm{d}\sigma^{\mathrm{IP}}_{xy}}{\mu_0\mathrm{d}H_x} \approx \SI{1}{((V/A)/T)}$, which is greatly improved compare to our previous work $(S_x = \SI{0.02}{((V/A)/T)})$~\cite{koraltan2023single}. Similar to OOP sensing, we investigated the sensing signals as a function of injected in-plane currents at room temperature and at $T=\SI{200}{K}$ (Extended Data Fig.~\ref{fig:EDFig07}), demonstrating once again the robustness of using the proposed skyrmionic device to generate a sensing signal.

Generally, the IP sensitive direction is that direction that is parallel to the applied current. Thus, it is to also sense $H_y$ fields simply by injecting the current along this direction. As this is equivalent to the case we discussed above due to simple symmetry arguments, we will not be providing an experimental demonstration.

\subsection*{\label{sec:level3}Sensing Performance}
\begin{figure}[]
    \centering
    \includegraphics[width=\columnwidth]{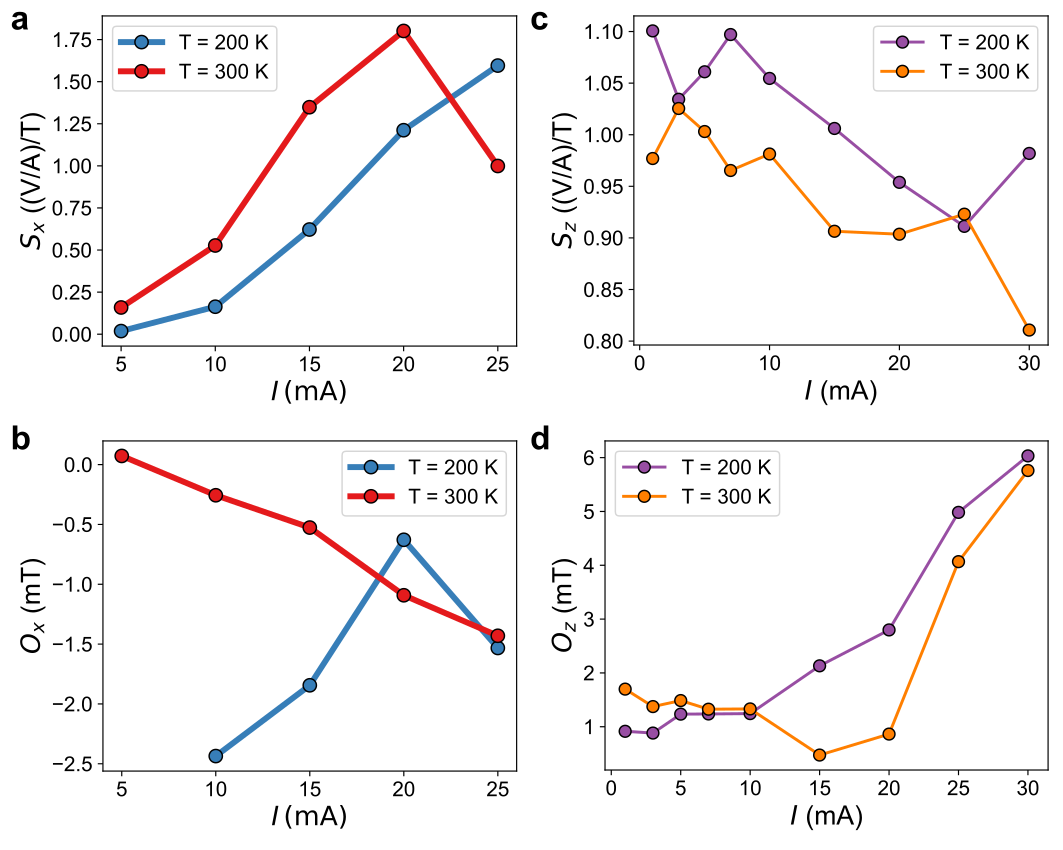}
    \caption{\textbf{Sensor performance at different environmental temperatures.} The calculated sensitivities for IP fields (a) and OOP fields (c) as a function of injected SOT current for both $T=\SI{300}{K}$ (red) and $T=\SI{200}{K}$ (blue) as environmental temperatures set in the cryostat. The slopes of the sensor transfer curves were fitted by a linear function to extract the sensitivities. The DC offset was then extracted from the transfer curves as the magnetic IP (b) and OOP (d) fields where the measured signal vanishes.}
    \label{fig:fig03}
\end{figure}

Magnetic field sensors can be classified based on basic properties such as measurable linear range, sensitivity, noise, offset and detectability. To better highlight the performance of our skyrmionic device as a magnetic field sensor, consider now the sensor signals from Extended Data Figs.~\ref{fig:EDFig06} and \ref{fig:EDFig07}, from which one can quantify the reachable sensitivities and DC offsets, as summarized in Fig.~\ref{fig:fig03}. 
Figure~\ref{fig:fig03}a depicts the sensitivity of our sensor for an IP field ($H_x$) at $T=\SI{300}{K}$ and $T=\SI{200}{K}$. The sensitivity can be increased by injecting more current into the skyrmionic device, until we reach a peak sensitivity of $S_x = \SI{1.85}{(V/A)/T}$ for $I=\SI{20}{mA}$. For an OOP field, the sensitivity is not affected as strongly as for IP fields; as we show in Fig.~\ref{fig:fig03}c. Here, the sensitivity is in the range of $S_z = \SI{0.95}{(V/A)/T}$, and it deviates only from $\pm 10 \%$ for higher and lower currents, respectively. Compared to commercially available TMR sensors, the sensitivities we reach are orders of magnitude lower, but we need to keep in mind that we read out the signal via the AHE, which is generally a much smaller effect. If one reads out the change in magnetization via magnetic tunnel junctions (MTJ)~\cite{guang2023electrical}, one could achieve resistivities as high as those of commercially available products. We demonstrate in Extended Data Fig.~\ref{fig:EDFig08} that we can enhance our sensitivity to IP fields by a factor of 4 simply by reducing the thickness of MgO to $\SI{1.1}{nm}$.

In a previous work~\cite{koraltan2023single}, in which we demonstrated the first offset-free sensing principle enabled by SOT with a linear range higher than $\SI{10}{mT}$, the sensitivities reached were as low as $\SI{0.017}{(V/A)/T}$. Thus, the skyrmionic device that we realized in this work increases the sensitivity by two orders of magnitude.
In the work of Li et al.~\cite{li2021spin}, they reach sensitivities of the order of $\SI{280}{(V/A)/T}$ for IP fields and one order of magnitude higher for OOP fields. However, their linear range is limited to only $\SI{\pm 1}{mT}$ (IP) and $\SI{\pm 0.4}{mT}$ (OOP), respectively. Similarly, AHE based sensors report very high sensitivities~\cite{zhu2014giant, peng2019ultrasensitive} up to three orders of magnitude higher than our sensor. While their linear ranges are again very limited ($< \SI{4}{mT}$), the thickness of the total magnetic active component is lower. As the AHE decreases with increasing thickness of the magnetic layer~\cite{grigoryan2017anomalous}, lower resistances are recorded in our skyrmionic sensor. Fortunately, this opens also new possibilities for skyrmion-based sensors of the future, where thin-film skyrmionic systems can be used to sense magnetic fields with much higher sensitivities and potentially even higher linear ranges.

As explained in more detail in the Methods section, we use the Quantum Design Physical Property Measurement System (PPMS), where the magnetic field is applied by a superconducting magnet. For a superconducting magnet, the applied field is linear to the sent current; thus, a very precise calculation of the magnetic field can be reached internally. However, we performed a linearity test by measuring the electron-paramagnetic resonance of 2,2-diphenyl-1-picrylhydrazyl (DPPH) to calibrate the applied field versus the expected resonance field. We find that the PPMS field has an offset of $\SI{-0.59}{mT}$ and a linearity error of $1.4\%$. This error becomes significant for very high magnetic fields, which we are not considering. However, the zero-field offset will, of course, limit the DC offset of our sensors. Without correcting this error, we provide in Fig.~\ref{fig:fig03}b ($O_x$) and in Fig.~\ref{fig:fig03}d ($O_z$) the DC offsets of our magnetic field sensor, which are the magnetic field for which $\sigma_{xy} = 0$. In the IP field, where we applied the differential measurement scheme and the offset-free sensing principle, very low offsets are recorded, $<\SI{1.5}{mT}$ at $T=\SI{300}{K}$. When the field of the PPMS is corrected, this error is reduced to $<\SI{1.0}{mT}$ and disappears almost completely for $I=\SI{10}{mA}$ and $I=\SI{15}{mA}$. For the OOP field, the situation changes significantly. Although the error is rather small at lower currents, as we simply measure the hysteresis curve of a conventional skyrmionic device, the DC offset error increases with the applied SOT current. Taking into account also the fact that we do not apply any type of magnetic field correction here, the measured offsets are rather high for the high current regime ($\SI{5}{mT}$ at $I=\SI{30}{mA}$). However, this error can also be eliminated by applying a spinning current concept~\cite{mosser2017spinning}. If the current is applied along $y$ instead of $x$, basically the same transverse voltage can be obtained, but mirrored with respect to the field if the contacts are chosen accordingly. Then, the two signals can be subtracted again from each other, leaving again an offset free signal behind for an OOP field as well.

\subsection*{\label{sec:level4}Origin of the sensing signal}
In the following we will take an in-depth look at the working mechanism of our 3D magnetic field sensor, where we make use of micromagnetic simulations at finite temperature using \texttt{magnum.np}~\cite{bruckner2023magnumnp} to analyze the field-dependent evolution of the magnetic states. 

\begin{figure*}[t]
    \centering
    \includegraphics[width=\textwidth]{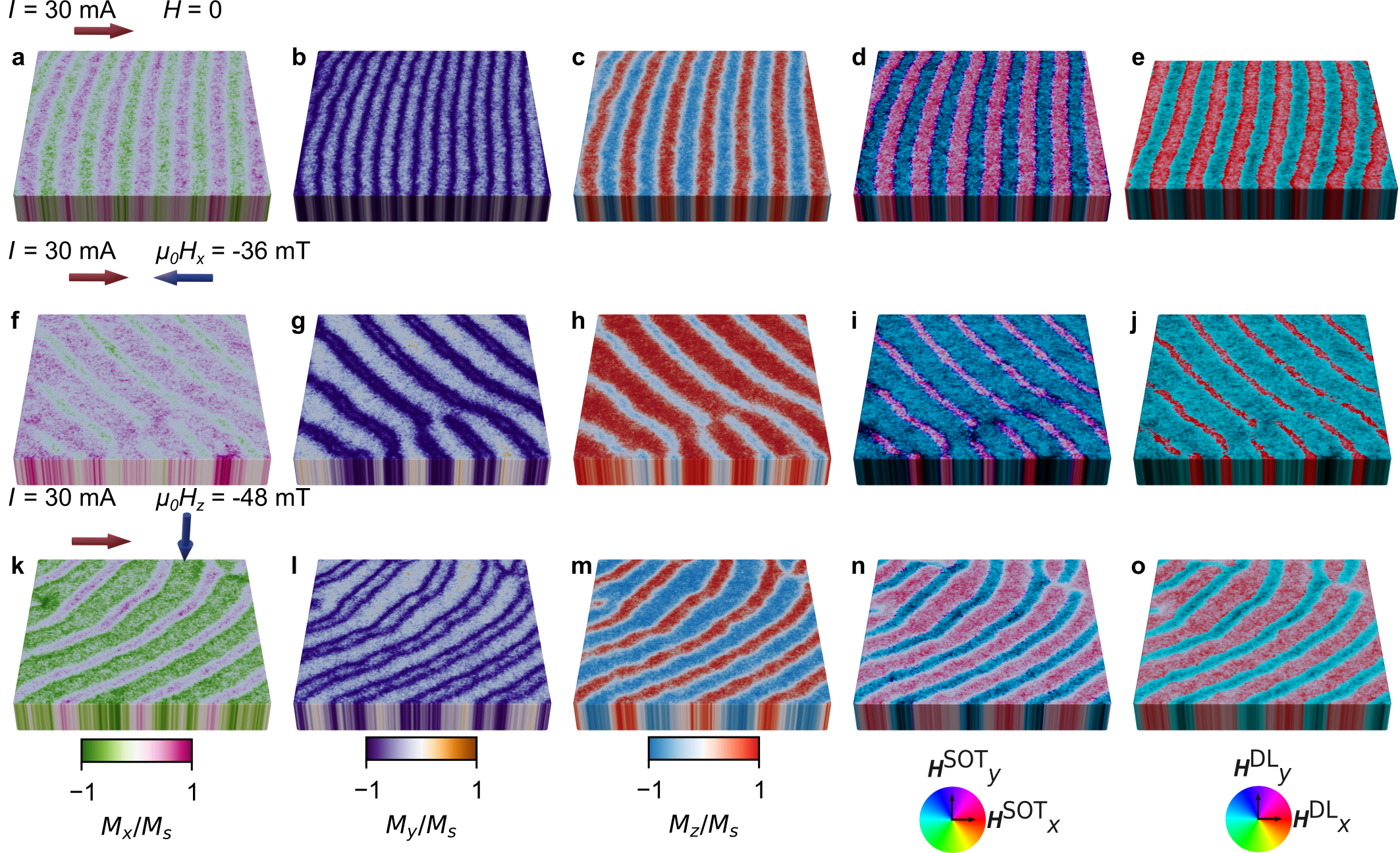}
    \caption{\textbf{Influence of effective spin-orbit torques on the nonuniform magnetization states.} Finite-temperature micromagnetic simulations demonstrating different scenarios for injected SOT current along $+x$ with a magnitude $I = \SI{30}{mA}$ and applied magnetic fields where each row represents a combination: (up) vanishing field, (center) IP field with $\mu_0H_x = \SI{-36}{mT}$, and (down) OOP field with $\mu_0H_z = \SI{-48}{mT}$}. The $x$, $y$ and $z$ components of the magnetizations are given in (a)-(c), (d)-(f), and (g)-(i), respectively. The effective normalized SOT field $\mathbf{H}^{\mathrm{SOT}}$ color coded in (j)-(l) by the color wheel at the bottom, as well as the normalized effective $\mathbf{H}^{\mathrm{DL}}$ to highlight the dominating effect of SOT in (m)-(o).
    \label{fig:fig04}
\end{figure*}

To understand the main mechanisms of SOT in combination of high Joule heating and magnetic fields, we studied three different cases with high SOT current of $I = \SI{30}{mA}$ in the $x$ direction. We start from a random magnetization state, turn on the current, and solve the LLG equation extended with the DLT and FLT from Eqs.~\eqref{eq:eq01} and \eqref{eq:eq02}. We include thermal fluctuations~\cite{leliaert2017adaptively} and temperature-dependent material parameters and numerically integrate the LLG until we reach an equilibrium state. No external magnetic field is applied. For the SOT coefficients, we assume $\eta_{\mathrm{DL}} = -0.3$ and $\eta_{\mathrm{FL}} = 0.05$ as appropriate values for the SOT efficiencies in W/CoFeB/MgO trilayers~\cite{manchon2019current}. It is worth mentioning that these are then reduced by the ratio of thickness of magnetic and nonmagnetic layers as a part of our effective media modeling; see Methods for more details.

The magnetization components are shown in Fig.~\ref{fig:fig04}a, b, c, where a well-ordered stripe pattern is visible, where the domain walls are all oriented along $-y$. Figure~\ref{fig:fig04}d illustrates the normalized direction of the total effective SOT field that acts on the magnetic layer, while Fig.~\ref{fig:fig04}e depicts only the normalized $\mathbf{H}^\mathrm{DL}$.  We do not illustrate $\mathbf{H}^\mathrm{FL}$ as this is a constant field along the polarization $\mathbf{p}$. 
Let us consider the expressions for the effective SOT fields in Eqs.~\ref{eq:eq01} and Eq.\ref{eq:eq02}. $\mathbf{H}_\mathrm{FL}$ tries to align the magnetization along the direction of the polarization, as an external field, while the direction of $\mathbf{H}_\mathrm{DL}$ depends on the local magnetization vector.
Our numerical investigations reveal that the DLT is mainly responsible for the periodic alignment of the stripes that exist intrinsically in the sample at vanishing fields and SOT currents. The FLT then forces all the domain walls and stripes to align towards $-y$. Note that we do not include the Oersted fields directly in our calculations but assume that the FLT already is an effective FLT where the contributions of Oersted fields are considered, as this separation is usually not done in the quantification of SOT parameters. In vanishing fields, DLT similarly favors an alignment along $-y$ if the magnetization is initially oriented OOP. To better understand the role of each torque individually, we varied the strength of both SOT coefficients. The higher $|\eta_\mathrm{DL}|$, the lower the average width of the stripes.

After the stripe domain is formed, we now apply an IP field $\mu_0H_x = \SI{-36}{mT}$. We observe that the external field affects the chirality of the system induced by FLT and reorients the walls approximately $\SI{40}{^\circ}$ from the $y$ axis towards $x$; as depicted in Fig.~\ref{fig:fig04}b,e,h. The strength of the applied field changes the angle of the stripe domains as a consequence of an effective field from the external field and FLT. Normalized $\mathbf{H}_\mathrm{SOT}$ is illustrated in Fig.~\ref{fig:fig04}k, while $\mathbf{H}_\mathrm{DL}$ is shown in Fig.~\ref{fig:fig04}n. Since the external field introduces a stronger $x$ component of magnetization, the DLT then is completed by an additional term along the $z$ component due to the cross-product $\mathbf{m}\times\mathbf{p}$. Furthermore, the DLT acting on the previously negative domains is enhanced, transforming them more IP, while the torque acting on the positive domains is minimized. Thus, the negative (down) domains shrink, while the positive (up) ones grow. Note that this effect is fully reversible with respect to the injected current, which means that, for a negative current and negative IP fields, the negative OOP domains grow, while the positive ones shrink, as $H_{\mathrm{DL},z}$ is now negative.
The magnetization state behaves differently for an OOP field $\mu_0H_z = \SI{-48}{mT}$, increasing the effect of the DL torque acting on the regions originally magnetized along $-x$ that favors a reorientation of the stripes along $+x$. Thus, the induced $x$ component of the magnetization will generate an effective OOP field, where effective field of FLT and the Zeeman term then additionally shrink the domains, as the new direction is tilted increasingly towards $-z$. Reversing the current polarity will reorient the stripes rather along $-x$, but the applied negative field will still lead to the shrinking of the positive domains due to the induced OOP SOT field.

\begin{figure*}[]
    \centering
    \includegraphics[width=\textwidth]{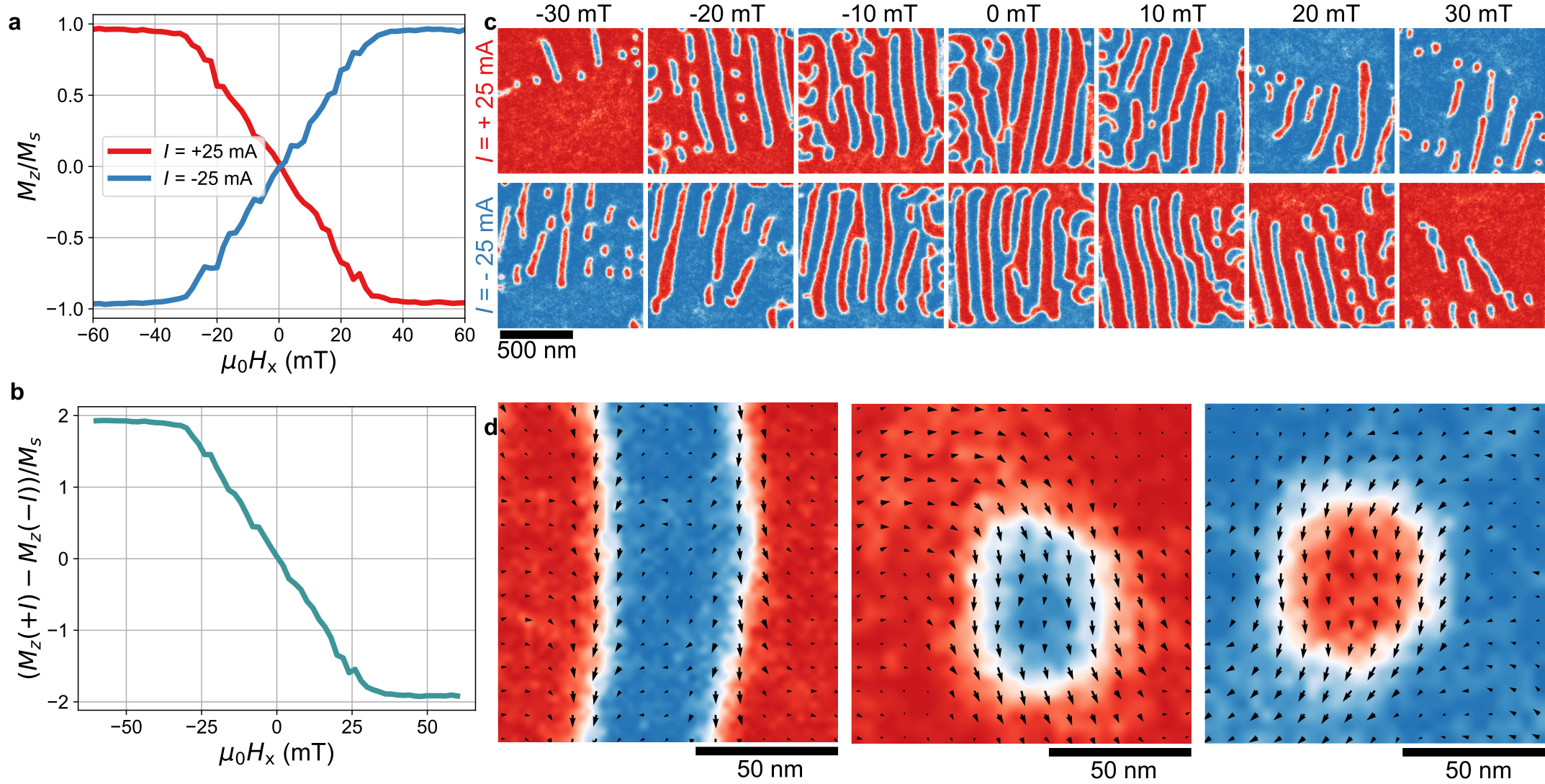}
    \caption{\textbf{Origin of sensing signal for IP fields.} (a) Micromagnetically simulated transfer curves for IP magnetic fields equivalent to Fig.~\ref{fig:fig02}e, and the calculated linear sensor signal (b) as the difference of the two curves from (a). Snapshots of the magnetizations (c) for positive current (top row) and negative current (bottom row) for magnetic fields in the range $\SI{\pm 30}{mT}$. Selected enlarged view of a stripe (left) at zero field, type-II bubbles with a negative core (center) and a positive core (right) in (d) at $\mu_0H_x = \SI{26}{mT}$ and $\mu_0H_x = \SI{-28}{mT}$, respectively. The arrows in (d) show the in-plane component of the magnetization.}
    \label{fig:fig05}
\end{figure*}

As the magnetization changes significantly for these fields, we obtain a transverse voltage dominated by the AHE, which we use to record a sensor signal in our experiments. Note that the AHE is proportional to $M_z$ in the micromagnetic simulations. Figure~\ref{fig:fig05} summarizes the results of our micromagnetic simulations for the IP sensing mechanism. The numerically obtained magnetization responses to the applied IP magnetic field are given in Fig.~\ref{fig:fig05}a, while the calculated sensor transfer curve can be seen in Fig.~\ref{fig:fig05}b. The applied current is reduced to $I = \SI{\pm 25}{mA}$. We have exceptionally good qualitative agreement with our experiments; thus, we are confident that we can make use of the underlying microstates to reveal the origin of the sensing signal. In Fig.~\ref{fig:fig05}c,d we provide snapshots of the magnetization states at different IP fields for both current directions. At vanishing fields, both currents lead to a similar stripe pattern. As the current is reduced compared to Fig.~\ref{fig:fig04}, the stripe domains are not as well structured as in the previous case. The domain walls around a stripe domain favor a parallel alignment with the FLT ($-y$ axis), as can be seen in Fig.~\ref{fig:fig05}d-left. In the upper row we provide the magnetization states for a positive current, where we observe that a positive field slightly reorients the stripes. The DLT favors negative domains with increasing $+x$ component of the magnetization due to cross-product $\mathbf{m}\times\mathbf{p}$. Therefore, the positive stripes break down into what looks like skyrmionic textures (type-II trivial bubbles). The vector field depicted in Fig.~\ref{fig:fig05}d shows that the FLT and DLT destroy the topological protection of the skyrmions, and rather topologically trivial type-II bubbles are observed~\cite{goebel2021beyond}. Note that skyrmionic states with an integer topological number would be moved out of the sample because of strong SOT currents~\cite{tomasello2014strategy}. The type-II bubbles are stable because their boundaries are aligned parallel to the polarization of the spin current; thus, the resulting torque vanishes. Due to the symmetry of the SOT, reversing the current polarity allows the DLT to favor positive domains, thus allowing the negative stripe to shrink down and ultimately collapse into type-II bubble states. Thus, a positive ferromagnetic state can be achieved by injecting a negative current and positive field, or a positive current and negative field, and a negatively polarized state by a negative current and negative field, or a positive current and positive field. Overall, the IP field enhances or diminishes the effect of DLT on positive or negative magnetized domains, based on the applied current direction. This in exchange leads to a transverse voltage due to the collapse of stripe domains in to skyrmionic states before fully saturating.

\begin{figure*}[]
    \centering
    \includegraphics[width=\textwidth]{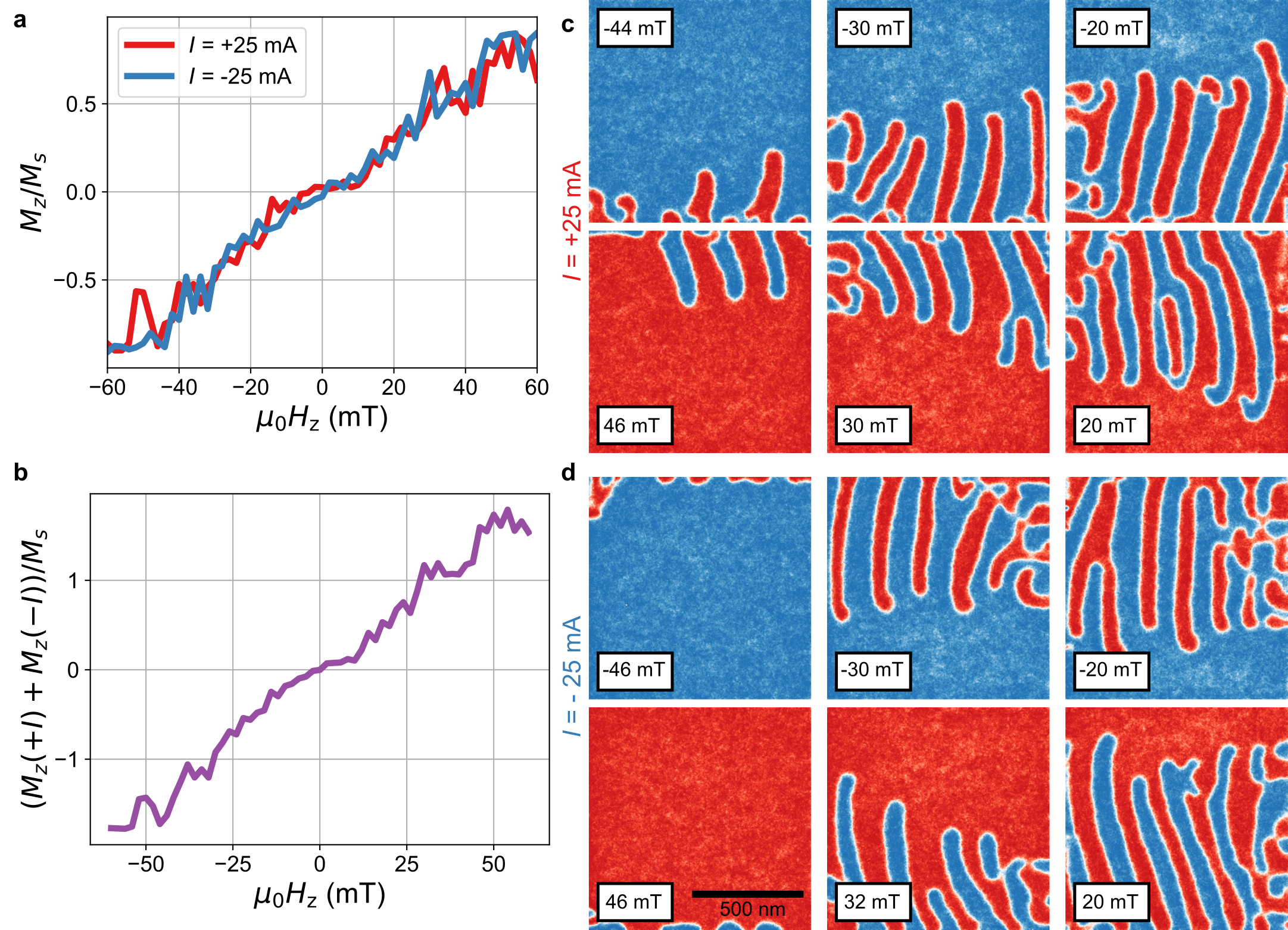}
    \caption{\textbf{Domain morphology during OOP field sensing.} Analogously to Fig.~\ref{fig:fig05}, we provide the transfer curves for positive and negative injected SOT currents in (a), and the sensor signal in (b) as the average of two curves from (a) as obtained from finite-temperature micromagnetics. The domain morphology is illustrated as snapshots of magnetizations for positive currents (c) and negative currents (d). The zero-field magnetization profiles are similar to those from Fig.~\ref{fig:fig05}, thus, we do not repeat them here. With increasing (decreasing) magnetic field, stripes with negative (positive) domains start to shrink, approaching a positive (negative) polarized ferromagnetic state.}
    \label{fig:fig06}
\end{figure*}

The case of an OOP magnetic field is much simpler. At vanishing fields, we find the same periodic stripe domain state with the orientation of the domain walls being dictated by the SOT. The simulated magnetization responses to the OOP field are depicted in Fig.~\ref{fig:fig06}a, and the sensor signal calculated from these two curves is shown in Fig.~\ref{fig:fig06}b. As described above, an OOP field slightly reorients the stripes along the $\pm x$ axis, which is also visible in the snapshots provided in Fig.~\ref{fig:fig06}c (positive current) and in Fig.~\ref{fig:fig06}d (negative current). Our finite-temperature micromagnetics reveal that with increasing (decreasing) OOP field positive (negative) domains are favored, while the system thrives to reach a saturated state. As the SOT induced a well-defined directionality and chirality of the stripe domains, the transformation of stripe domains into the polarized states occurs hysteresis-free, leading to a highly linear sensor signal.

\section*{\label{sec:level5}Discussion and Conclusions}
In this work, we demonstrated that we can use a skyrmionic device to sense both IP and OOP magnetic fields with linear ranges up to $\SI{\pm 17}{mT}$ and $\SI{\pm 30}{mT}$, respectively. The applied fields will either act with or against the damping-like torque and lead to the collapse of periodically arranged stripes into skyrmionic states (type-II bubbles) of given core polarity, as we reveal employing finite-temperature micromagnetic simulations. The change in the magnetization can be measured as a change in the transverse voltage, where the main contributor is the anomalous Hall effect. Repetition of our experiments also at lower temperatures ($T=\SI{200}{K}$) shows that our skyrmionic sensor could potentially also be used at cryogenic conditions. As the main objective, we demonstrated offset cancelation, where the DC offset can be nearly eliminated for IP magnetic fields by employing an offset-free sensing principle, where the symmetry of the SOTs is exploited. The offset present for the OOP fields can eventually be eliminated by applying a spining-current technique based on the symmetry of the interactions. The proposed skyrmionic device works very well for larger systems, where we have enough stripes to transform to type-II bubbles and to the polarized state. Furthermore, the domain wall width as well as the pinning fields of the samples will significantly impact the sensing performance.

The sensitivity is orders of magnitude lower when compared to commercially available TMR sensors, as well as AHE-based sensors with ultrathin ferromagnetic layers. Several improvements can be made to improve the sensitivity. We have demonstrated here that one can reduce the thickness of the MgO layer to further increase the resistivity of our rather thick multilayer system and hence ameliorate the sensitivity. A different option is to change the material of the skyrmionic device. A multilayer skyrmionic stack composed of [Ir / Fe / Co / Pt]$_{\mathrm{N}}$  can be engineered to host a high density of skyrmions and stripe domains~\cite{soumyanarayanan2017tunable}. As these stacks are much thinner, the measured resistances are expected to increase accordingly. Because of the Pt and Ir interfaces, very strong SOTs are expected in such a stack. While the thickness reduction improves the sensitivity, the high density of skyrmions/stripes and stronger torques are expected to increase the linear ranges to the levels of commercially available TMR sensors. Furthermore, stronger effective SOT fields can be reached within such a stack due to additive spin current generation, which will further increase the linear range. Ultimately, one could combine the present approach (or [Ir/Co/Fe/Pt]$_{\mathrm{N}}$ based stack) with the read-out of the OOP component of the magnetization by using a magnetic tunnel junction and exploiting the TMR effect~\cite{guang2023electrical}, specifically enhancing the sensitivity, while keeping the DC offset minimal. To do so, the MTJ needs to consist of an OOP syntethic antiferromagnet, for instance, Co/Pt/X/Pt/Co multilayers, where X can be Ir, Ru or a similar material that ensures a high interlayer exchange interaction~\cite{fernandez2019symmetry}.

Our results represent an innovative approach to combine both skyrmions and SOT based spintronic devices for magnetic field sensing applications. Our experimental and numerical investigations will pave the way towards a new research direction for magnetic skyrmions and magnets with nonuniform magnetization states to be explored as potential candidates for magnetic field sensing applications.

\section*{\label{sec:level6}Methods}

\textbf{Sample Growth and Patterning}\\
\textit{Thin films}. For the characterization of magnetic parameters the [W(5)/CoFeB(0.7)/MgO(1.2)]$_{\mathrm{N}}/\mathrm{HfO_2}$ multilayer stack was sputter deposited on an 8-in silicon wafer using a Singulus Rotaris tool deposition
system with base pressure $<\SI{3e-8}{mbar}$ using thermally oxidized silicon substrates. The stack is then capped with $\mathrm{HfO_2}(3)$ to avoid oxidation and prevent electromagnetic coupling to the coplanar waveguide for FMR measurements. Metals were deposited in the DC mode, while oxides ($\mathrm{MgO},\, \mathrm{HfO_2}$) were deposited in the RF mode.\\
\textit{Hall Bars}.
We apply the same strategy as in \cite{koraltan2023single}. That is, before film deposition, the wafers received an aluminum metallization layer to realize the contacts for current flow into the stack from the bottom, where W vias are used through an insulation $\mathrm{SiO_2}$ layer. A mechanical chemical polish is then carried out to ensure a smooth surface for the deposition of the [W(5)/CoFeB(0.7)/MgO(1.2)]$_{\mathrm{N}}/\mathrm{Ta}(3)$ stack. We use the Singulus Rotaris tool to sputter-deposit the multilayer stack on the prepared substrate, as described above for thin films. The structure is then capped with Ta(3) to avoid oxidation. Hall bar patterning is achieved using conventional optical lithography and $\mathrm{Ar^+}$ ion etching. The average milling rate for the SOT stack was found to be around $\SI{10}{nm/min}$.

\textbf{Vibrating Sample Magnetometry}. For experimental investigations, we use the Physical Property Measurement System (PPMS) from Quantum Design where magnetic fields up to $\SI{\pm 9}{T}$ can be applied, while the temperature can be varied between $\SI{1.8}{K}-\SI{400}{K}$. In order to quantify the magnetic moment as a function of the applied magnetic field, we use the Vibrating Sample Magnetometer module. For this, we mechanically cut the thin film specimen to approximately $2.5\times 2.5\, \mathrm{mm^2}$, in order to measure the hysteresis curves while applying the field both parallel (IP) and normal (OOP) to the film plane. The oscillation amplitude for the VSM is chosen as $\SI{2}{mm}$, while the frequency is $\SI{40}{Hz}$. We measured the hysteresis loops for both configurations for temperatures between $\SI{25}{K}-\SI{400}{K}$ in $\SI{25}{K}$ steps, as provided in Extended Data Fig.~\ref{fig:EDFig01}. As we measure the total magnetic moment of the sample in $\mathrm{Am^2}$, we then use the magnetic volume to obtain the saturation magnetization $M_s$ as a function of temperature. We found $M_s = 1375.25\times \left(1-\left(T/T_C\right)^{1.77} \right)\, \mathrm{(kA/m)}$, where $T_C$ is the Curie temperature of the magnet. The fit yields $T_C = \SI{952.63}{K}$. Note that we normalize $M_s$ to the magnetic volume only ($t_\mathrm{m} = \SI{7}{nm}$). For the micromagnetic simulations, we scale the magnetic moment to the total volume. We do this to employ an effective media model for our micromagnetic modeling, as discussed later.  In our case, we have a magnetic thickness $t_\mathrm{m} = \SI{7}{nm}$ and a nonmangetic thickness of $t_\mathrm{nm} = 10t_{W} + 10t_{MgO} = \SI{62}{nm}$. Thus, for the effective media model's saturation magnetization, we obtain $M^{\mathrm{emm}}_s = 139.52\times \left(1-\left(T/T_C\right)^{1.77} \right)\, \mathrm{(kA/m)}$ as the best fit.

\textbf{Ferromagnetic Resonance}
We perform ferromagnetic resonance (FMR) measurements to quantify the perpendicular magnetic anisotropy in our samples. For this purpose, we use a sample-holder with a coplanar waveguide (NanOsc) such that the magnetic field applied within the cryostat is IP, where we employ the flip-chip method to investigate the field dependence of the resonance frequencies. To do so, we use a Rohde\&Schwarz Z40A Vector Network Analyzer to apply and detect microwave signals with frequencies up to $\SI{40}{GHz}$. The signal power is $\SI{5}{dBm}$, to obtain high signal-to-noise ratios and well-defined resonance peaks. We then measure the scattering parameter $S_{21}$ in a frequency sweep mode in a fixed magnetic field. The obtained resonance frequencies $f_\mathrm{res}$ of the fundamental peak are evaluated against the resonance fields $H_\mathrm{res}$. For an IP field configuration the resonance condition is given by the Kittel formula $f_\mathrm{res} = \gamma' \mu_0 \sqrt{H_\mathrm{res}(M_\mathrm{eff}+H_\mathrm{res}{})}$, where $\gamma'$ is the reduced gyromagneitc ratio, $\mu_0$ is the vacuum permeability, and $M_\mathrm{eff}$ is the effective magnetization, which is given by $M_\mathrm{eff} =  M_s\left( 1 - \dfrac{2K_u}{\mu_0M_s^2}\right)$,
with $K_u$ being the uniaxial anisotropy constant. In our case, this equals the strength of the perpendicular magnetic anisotropy originating from the heavy metal feromagnet interface. We repeat the measurements for the same temperature range as the VSM measurements. Since $M_s$ is known, we use the Kittel resonance condition to fit both the reduced gyromagnetic ratio and $M_\mathrm{eff}$ using  least squares methods. The average reduced gyromagnetic ratio is found to be $\gamma' = \SI{27.66}{GHz/T}\pm \SI{0.27}{GHz/T}$. In Extended Data Fig.~\ref{fig:EDFig02} we display the measurement setup, as well as the extracted temperature dependent quantities $K_u$, $M_\mathrm{eff}$, $K_\mathrm{eff}$, which denotes the effective anisotropy constant, $H_{k, \mathrm{eff}}$ which is the effective anisotropy field. The temperature dependence of $K_u$ is found to be $K^\mathrm{emm}_u (T) = 1.43\times (M_s(T)/M_s(0))^{2.28}\, \mathrm{(MJ/m^3)}$, if one considers $M_s$ normalized to magnetic layers only. For micromagnetic simulations, we use $M^\mathrm{emm}_s$ for which we obtain $K^\mathrm{emm}_u (T) = 36.7\times (M_s(T)/M_s(0))^{3.21}\, \mathrm{(kJ/m^3)}$.

\textbf{Magnetotransport Measurements.}
Patterned 6-arm Hall bars are used to electrically investigate the feasibility of the skyrmionic device as a magnetic field sensor. For this purpose, four contacts are established by wire-bonding $\SI{25}{\mu m}$ gold wires. This enables to measure the transverse voltage through a four-point measurement, whereas the longitudinal resistance $R_{xx}$ is determined in two-point geometry at the current contacts. We use a self-built box to control the channels along which the current is applied. The current, always injected along the x axis, is delivered by a Keithley 6221 current source, which can apply currents up to \SI{100}{mA}. We use a Keithley nanovoltmeter (model K2182) for voltage measurements, which has two channels and offers nV resolution. With the first channel, we record the transverse resistance $R_{xy}$, with the second channel the longitudinal resistance $R_{xx}$.
In order to apply IP and OOP fields during measurements, we mount the sample in a horizontal rotator, which allows to control the angle of the magnetic field by a linear motor.  The sample holder and the chamber are continuously grounded while the sample is inserted into the PPMS.
For all experiments, we start from the highest positive magnetic field and apply a differential measurement scheme. That is, we consequently measure the resistances for positive and negative currents (three times each, and then we take the average for each current direction). Afterwards, the magnetic field is lowered and we repeat this procedure until the lowest magnetic field has been reached.
As we are using the horizontal rotator, small angle deviations must be expected and might disturb the sensing signal, leading to poor performance. However, our results indicate that we have successfully omitted large errors. As mentioned in the main text, we have tested the calibration of the superconducting magnet. We found that we have an offset error of $\SI{-0.59}{mT}$ and a linearity error of 1.4. To be more transparent, we have not corrected the magnetic fields in our post-processing, as we focus rather on the concept we are demonstrating.

\textbf{Magnetic Force Microscopy}
The MFM measurements presented in this study were performed using a home-built high vacuum $(\approx \SI{e-6}{mbar})$ scanning probe microscopy system where magnetic fields as high as $\SI{300}{mT}$ can be applied~\cite{feng2022quantitative}. The MFM is operated in vacuum, drastically improving the mechanical quality factor $Q$ for the cantiveler to $Q\approx\SI{2e5}{}$. In addition to this improvement over conventional MFM measurements in air, which increases the sensitivity by a factor of 40, we also deposit a very thin Co layer on the tip to minimize stray-field interactions with the sample. We choose SS-ISC cantilevers from Team Nanotech GmbH, where the tip radius is below $\SI{5}{nm}$. To enhance the sensitivity of the cantilever tip to magnetic fields, we utilized sputter deposition to apply a 3 nm layer of Co at room temperature on top of a Ta seed layer (2 nm), followed by capping it with a 4 nm layer of Ta to protect against oxidation. The oscillation of the cantilever at resonance with a constant amplitude of 7 nm and the detection of frequency shifts resulting from the derivative of the tip-sample interaction force were carried out using a Zurich Instruments phase-locked loop (PLL) system. It is important to note that the frequency shift indicates an attractive force (derivative) when it is negative. In Figs.\ref{fig:fig01} and \ref{fig:EDFig04}, if a negative OOP field is applied, the MFM tip is magnetized down, while skyrmions (stripes) will have a positively magnetized core. As the force generated between the spin textures and the tip is repulsive, the recorded frequency shift is negative. Hence, we revert the colors of the MFM images based on the magnetization of the tip to depict the magnetization states similar to those obtained from micromagnetic simulations and to omit artificial confusions.

\textbf{Micromagnetic Simulations}\\
\textit{Finite-temperature micromagnetics}. We use \texttt{magnum.np}~\cite{bruckner2023magnumnp} for GPU-accelerated micromagnetic simulations using a finite-difference algorithm for the numerical solution of the Landau-Lifshitz-Gilbert equation~\cite{abert2019micromagnetics} which describes the temporal evolution of the magnetization, where

\begin{equation}
\frac{\mathrm{d}\mathbf{m}}{\mathrm{d}t} = -\gamma \mathbf{m} \times \mathbf{H}^{\mathrm{eff}} + \alpha \mathbf{m} \times \frac{\mathrm{d}\mathbf{m}}{\mathrm{d}t},
\label{eq:llg}
\end{equation}

with $\alpha$ being the Gilbert damping parameter, and $\mathbf{H}^{\mathrm{eff}}$ the effective field term that is derived from the total energy of the system via the variational derivative as

\begin{equation}
\mathbf{H}^{\mathrm{eff}} = -\dfrac{1}{\mu_0M_s}\dfrac{\delta E_{tot}}{\delta\mathbf{m}},
\label{eq:llg}
\end{equation}
where $E_{\mathrm{tot}}$ denotes the total energy of the system. In our simulations, the energy contributions considered are the demagnetizing energy, exchange energy, perpendicular magnetic anisotropy energy, and DMI energy. All simulations are performed at finite temperatures, where a stochastic term is added to introduce thermal fluctuations into the system, which is directly added to the effective field via the thermal field $\mathbf{H}^\mathrm{th} = \boldsymbol{\epsilon} \sqrt{\dfrac{2\alpha k_B T}{{\mu}_0 M_s \gamma V \Delta t}}$, where $\boldsymbol{\epsilon}$ is a random vector that is normally distributed for each timestep $\Delta t$, $V$ is the cell volume, $k_B$ is the Boltzmann constant and $T$ is the temperature of the system. The now stochastic LLG is integrated using a Runge-Kutta-Fehlberg algorithm with an adaptive time step~\cite{leliaert2017adaptively}. Note that the results can depend on the mesh size, and a rescaling of parameters can improve the reproducibility of the results~\cite{Oezelt2022}. To reduce computation times, we used an effective media model, where it is assumed that the magnetization is homogeneous along the $z$ axis, and that all skyrmions and stripes are complete throughout the thickness. In this model, the effective media parameters are usually calculated from the parameters of a single magnetic layer using the relation $M_s = M_s^\prime\times t_\mathrm{m}/t_\mathrm{t_nm}$, where $M_s^\prime$ is the parameter of a single layer, $t_\mathrm{m}$ denotes the total thickness of the magnetic layers and $t_\mathrm{nm}$ the total thickness of the non-mangetic layers, respectively. However, we follow a different, yet equivalent, route to apply the effective media model. That is, we assume that the sample is magnetic throughout the thickness and that the measured magnetic moment from VSM results from the entire volume. Thus, we simulate the multilayer as a single-layer magnet with reduced saturation magnetization. The temperature dependence $M_s = 139.52\times \left(1-\left(T/T_C\right)^{1.78} \right)\, \mathrm{(kA/m)}$ is obtained, which is then used to fit the strength of the perpendicular magnetic anisotropy and its temperature dependence as $K_u (T) = 36.7\times (M_s(T)/M_s(0))^{3.2}\, \mathrm{(kJ/m^3)}$. The exchange constant $A^\prime_\mathrm{ex}$ for one W/CoFeB/MgO trilayer is taken from the literature as $A^\prime_\mathrm{ex} = \SI{15}{pJ/m}$. We assume that $A^\prime_\mathrm{ex}$ follows a dependence of $M_s$, where $A^\prime_\mathrm{ex}(T) = A^\prime_\mathrm{ex}(0)\times \left(\dfrac{M_s(T)}{M_s(0)}\right)^{1.7}$. This value is then reduced according to the effective media model as $A_\mathrm{ex}(T) = A^\prime_\mathrm{ex}(T)\times\dfrac{t_\mathrm{m}}{t_\mathrm{nm}}$. Note that all relevant material parameters ($M_s$, $K_u$, $A_\mathrm{ex}$ and $D$) are assumed as average values in our simulations. To better reproduce the reality, we assume that $M_s$, $A_\mathrm{ex}$ and $D$ are normally distributed around their mean values with a standard deviation of $20\%$, and $K_u$ with a standard deviation of $30\%$. Furthermore, as it is expected that the perpendicular anisotropy will not be perfect in the defects, we distribute the anisotropy axis in a cone around [001], where the maximal deviation angle is $\SI{30}{^\circ}$. \\
\textit{Numerical optimization of the DMI constant}. The value of the DMI constant at $T=\SI{0}{K}$ was optimized numerically as demonstrated in Fig.~\ref{fig:EDFig03}. For this purpose, we read the initial magnetization pattern from the MFM image of Fig.\ref{fig:fig01}k. The simulation box is discretized in $(1024\times 1024\times 1)$ cells, with total dimensions of $(\SI{4}{\mu m} \times \SI{4}{\mu m}\times \SI{69}{nm})$. The domain walls are then randomly magnetized in each cell, since we do not have any information about this from the MFM data. We relax the structure with moderately high damping $\alpha = 0.1$ for $\SI{10}{ns}$. The average absolute difference between the final relaxed state and the initial state is then calculated and plotted against the DMI value. We take the value at the local minimum as the optimal DMI value to reproduce our experiments, which is $D(0)=\SI{0.14}{mJ/m^2}$. For the remainder of the simulations, we will use this value, which depends on $M_s(T)$ as $D(T) = D_0\times (M_s(T)(M_s(0))^2$.\\
\textit{MH-Loops.}For the simulations in Fig.~\ref{fig:fig01} we use the same simulation box and parameters. For the hysteresis curve in Fig.~\ref{fig:fig01}b we always start from a randomly magnetized state, set an OOP magnetic field, and numerically solve the LLG for $\SI{10}{ns}$ for each field, including thermal fluctuations. Snapshots of the magnetizations in Fig.~\ref{fig:fig01} were obtained from a set of simulations, where we always start from the magnetization pattern that was measured by MFM, set an OOP magnetic field, and relax the system for $\SI{10}{ns}$.\\
\textit{Spin Orbit Torques simulations}. For these simulations, we reduce the total size of the simulation box to $(\SI{1}{\mu m}\times \SI{1}{\mu m}\times \SI{69}{nm})$ which is discretized in $(256\times 256\times 1)$ cells. The influence of the SOT current in the simulations is considered together with the effect of Joule heating. In our experiments, we quantified the sample temperature as a function of the applied lateral current, resulting in $T(I_x) = T_0\times(a\times I_x^2 + b\times I_x + c) - T_1$, where $T_0 = \SI{27.81}{K/\Omega}$, $a=\SI{9e-3}{\Omega/(mA)^2}$, $b = \SI{4.1e-3}{\Omega/(mA)}$, $c = \SI{68.31}{\Omega}$ and $T_1=\SI{1567.3}{K}$. Following again the effective media model approach, it is assumed that the current density that contributes to the spin polarization is achieved in individual layers, thus the current density employed in the effective SOT fields is calculated with the cross section $j_e = I/A = I/(\SI{1}{\mu m}\times \SI{6.9}{nm})$. Furthermore, the dimensionless SOT coefficients $\eta_\mathrm{dl}$ and $\eta_\mathrm{fl}$ are scaled according to the effective medium approach by $t_\mathrm{m}/t_\mathrm{nm}$. So in reality, one could keep the original cross section and assume that the total effective SOT parameters are unchanged. We tried to quantify the SOT parameters for our multilayer stack, but the signal obtained from second-harmonic measurements was too low to make a conclusive statement about the strength of the parameters. For the simulations presented in Figs.~\ref{fig:fig04}-\ref{fig:fig06} we always start from a random magnetization in each simulation cell. For each magnetic field,  we first apply the positive current and relax the system for $\SI{10}{ns}$, and then reverse the current polarity and relax for another $\SI{10}{ns}$. By doing so, we can apply the differential measurement protocol from the experiments.

\subsection*{Acknowledgements} S.K. thanks Barbora Budinska for support with PPMS measurements, Silvia Damerio, Alejandro de Souza, Stefano Fedel and Can Onur Avci for trying experiments of higher harmonic measurements for the quantification of SOT efficiencies, MOKE imaging, and fruitful discussions, and Wolfgang Lang and Andrii V. Chumak for the measurement equipment and use of their laboratories. We thank Maria-Andromachi Syskaki for the optimization of the HfOx growth process. The computational results presented have been achieved, in part, using the Vienna Scientific Cluster (VSC). S.K. and C.A. gratefully acknowledge the Austrian Science Fund (FWF) for support through Grant No. P34671 (Vladimir). S.K. and D.S. acknowledge the Austrian Science Fund (FWF) for support through Grant No. I 6267 (CHIRALSPIN). S.K. and D.S. acknowledge funding from {\"O}sterreichische Forschungsförderungsgesellschaft (FFG) under the project Senstronic. B.A. was supported by the Austrian Science Fund (FWF) through Grant No. I4865-N (FLUXPIN). Thin film deposition used infrastructure provided by the ForLab MagSens. R.G., F.K., I.K., G.J., and M.Kl. acknowledge support by the Deutsche Förderung Geselschaft (SFV TAR, 73 SPIN+x, A01, B02) and Infineon. The infrastructure for thin-film deposition was provided by the ForLab MagSens.

\subsection*{Author Contributions}
A.S. and D.S. conceived the project for a magnetic field sensor enabled by SOT. D.S. conceived the idea of using multilayers. S.K. conceived the idea of skyrmionic textures for sensing.
S.K. and B.A. built the experimental setup at the PPMS, S.K. performed all measurements (VSM, FMR, and transport measurements) and analyzed the data. S.K. performed and analyzed all micromagnetic simulations. R.G., F.K., I.K, G.J, M.Kl. sputter-deposited the thin films. R.P., A.O.M., and H.J.H. performed the MFM measurements, M.Ki. and K.P. fabricated the Hall bars,  S.H. and F.B. implemented the thermal fluctuations in the micromagnetic code, and S.K., F.B., C.A. and D.S. wrote and improved the micromagnetic code. D.S. supervised the project. S.K. wrote the initial manuscript. All authors have discussed the results, commented on, and improved the initial manuscript.  

\subsection*{Additional information}

\subsection*{Correspondence and requests for materials}
\bibliography{manuscript}

\newpage
\clearpage 
\onecolumngrid 

\setcounter{figure}{0} 
\setcounter{table}{0} 
\setcounter{equation}{0} 
\makeatletter 
\renewcommand{\thefigure}{ED\@arabic\c@figure}
\renewcommand{\thetable}{ED\@arabic\c@table}
\renewcommand{\theequation}{ED\@arabic\c@equation}
\makeatother

\begin{figure*}
    \centering
    \includegraphics[width=0.9\textwidth]{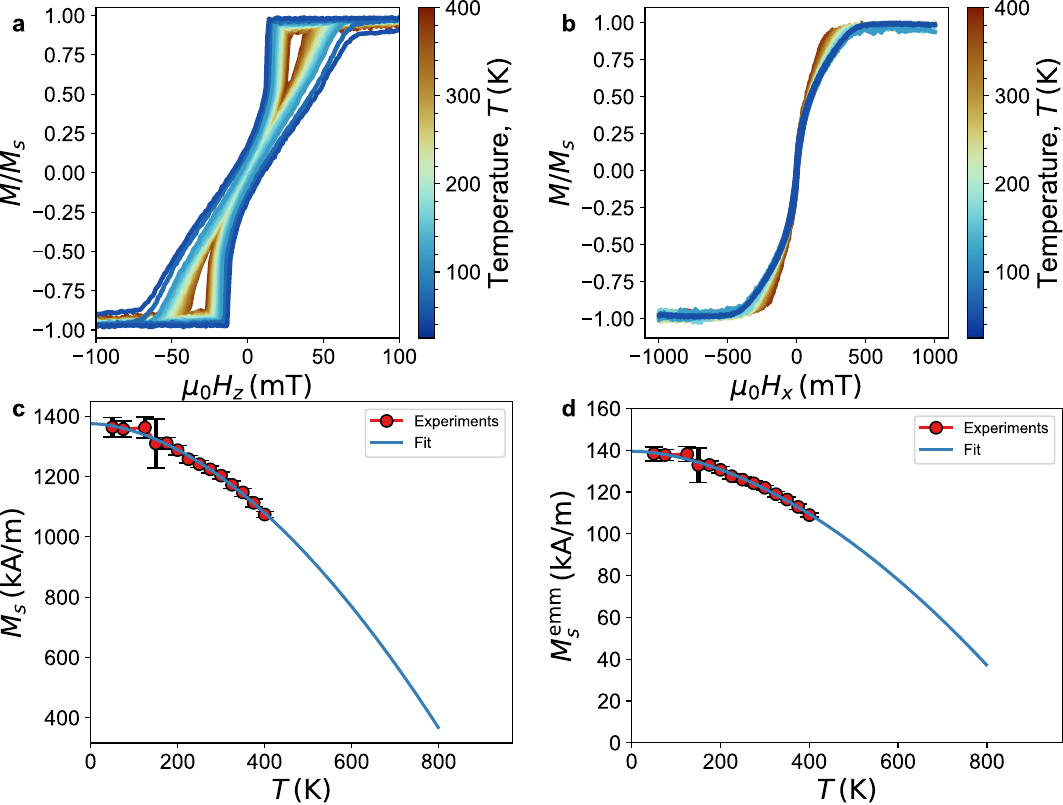}
    \caption{\textbf{Vibrating Sample Magnetomery measurements (experiments)}. Normalized MH-loops for OOP fields at different temperatures are given in (a), and for IP fields in (b). The saturation magnetization as a function of temperature was quantified in (c), where the magnetic moment was normalized to the magnetic volume only, while in (d) it was magnetized to the total volume according to the effective media approach.}
    \label{fig:EDFig01}
\end{figure*}

\begin{figure*}
    \centering
    \includegraphics[width=\textwidth]{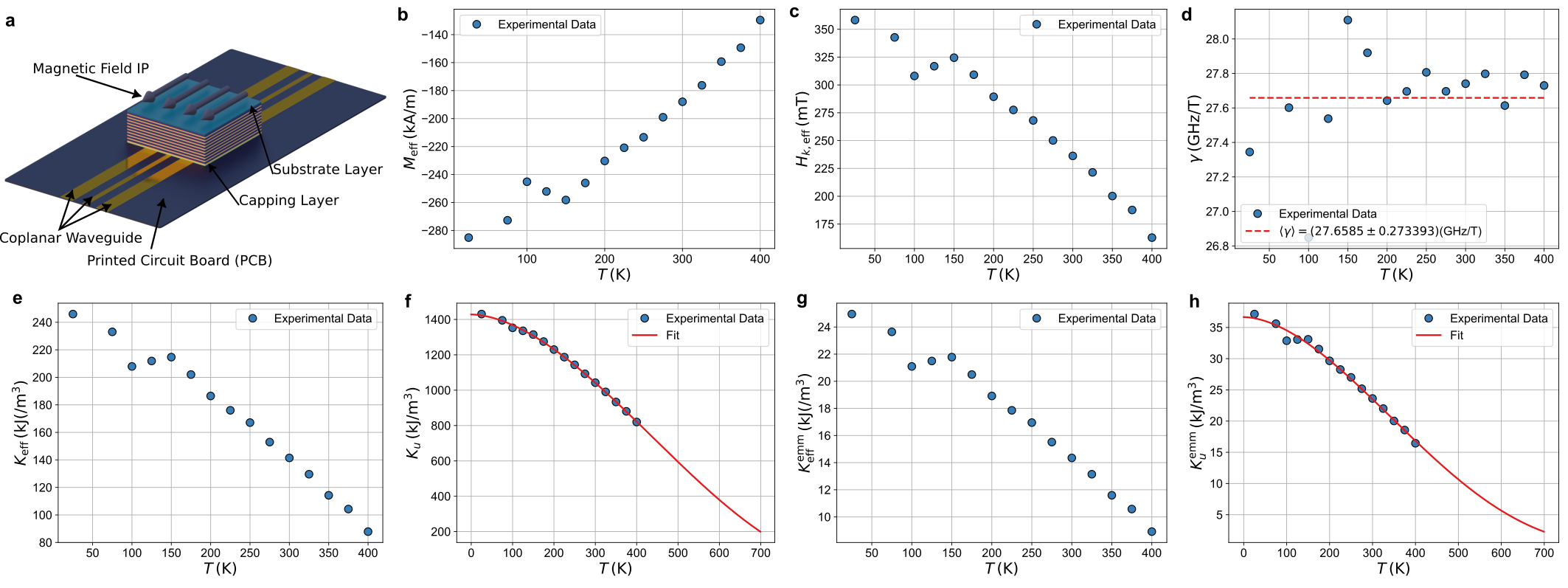}
    \caption{\textbf{Ferromagnetic Resonance Measurements (experiments)}. (a) Schematic illustration of the experimental setup, where we employ the flip-chip method, with the sample being placed in direct contact to a commercially available coplanar waveguide (CPW) on a printed circuit board (PCB). The magnetic field is applied IP, and the resonance frequencies are measured for fields where the sample is saturated IP. From the Kittel modes, we extract the effective magnetization $M_\mathrm{eff}$ as a function of temperature. For known values of the saturation magnetization, the effective anisotropy field shown in (c) is $H_\mathrm{k,eff} = \dfrac{2K_\mathrm{eff}}{\mu_0M_s}$, where the effective anisotropy constant is $K_\mathrm{eff} = K_u - K_d = -\dfrac{K_d\times M_\mathrm{eff}}{M_s}$, where $K_d$ is the shape anisotropy of the sample given by $\dfrac{1}{2}\mu_0M_s^2$. The reduced gyromagnetic ratio $\gamma'$ shown in (d) is also fitted from the Kittel modes. If the $M_s$ is normalized to the magnetic volume only we obtain $K_\mathrm{eff}$ and $K_u$ as illustrated in (e) and (f), respectively. If one uses the effective media model, the dependencies from (g) and (h) are obtained.}
    \label{fig:EDFig02}
\end{figure*}

\begin{figure*}
    \centering
    \includegraphics[width=0.9\textwidth]{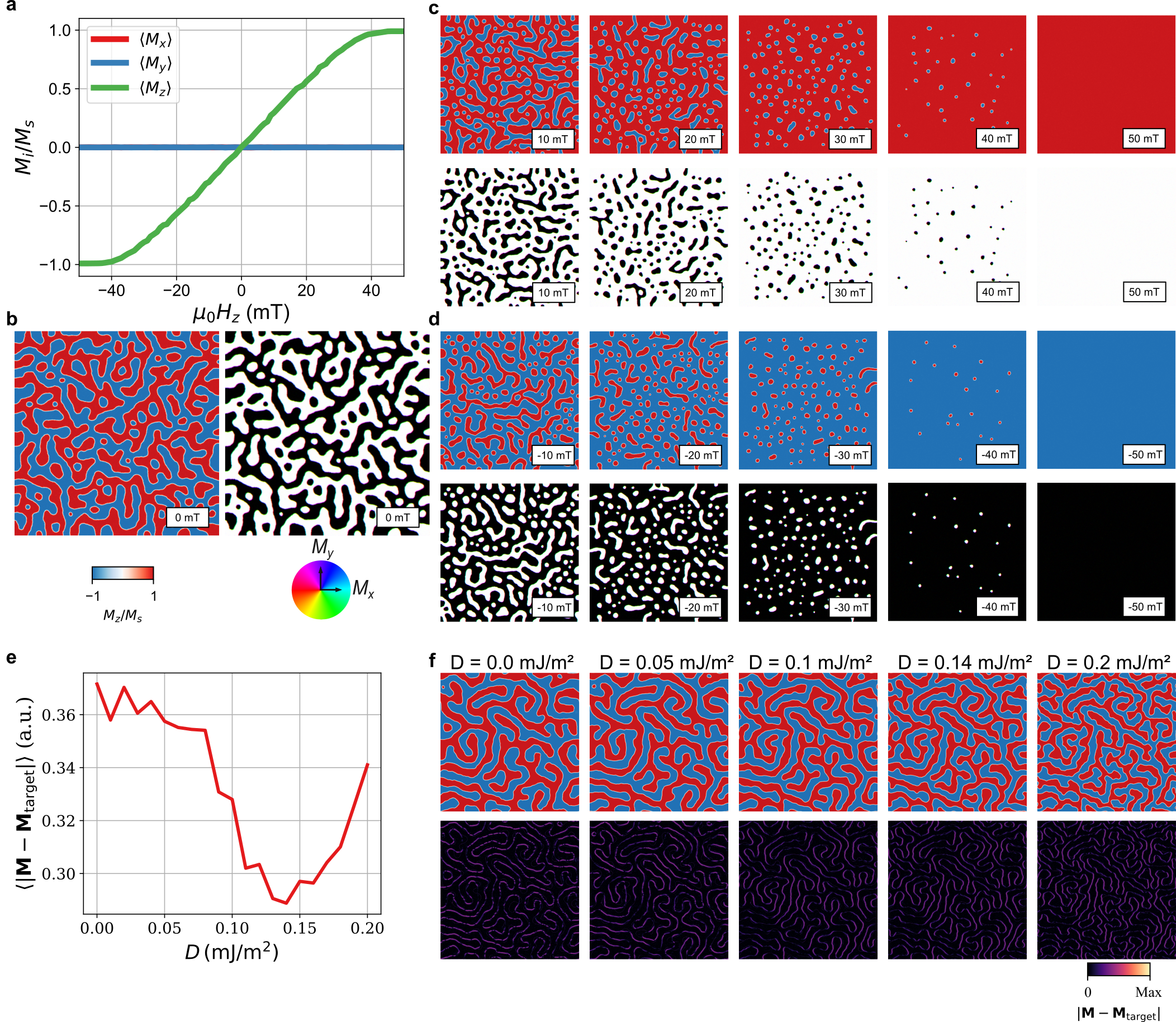}
    \caption{\textbf{Finite-temperatue micromagnetic and numerical optimization of DMI constant (simulations)}. Evolution of all components of the magnetization averaged spatially plotted as a function of the applied OOP field $\mu_0H_z$ in (a), while the zero field domain morhopology is displayed in (b) as the OOP component of the magnetization and with a Hue-Saturation-Value (HSV) coding scheme, where the color represents the orientation of IP components, while the dark ($-z$) and bright ($+z$) contrasts show the OOP components. Snapshots of the magnetization state for positive fields are given in (c) for positive OOP fields, and in (d) for negative OOP fields. Note that these magnetization states have been obtained by always starting from normalized random magnetization in each cell and then relaxing the system at each OOP field sequentially. The value of the DMI constant $D$ was numerically optimized where we provide in (e) the average value of the objective function $\langle |\mathbf{M} - \mathbf{M_\mathrm{target}}| \rangle$ that calculates the difference between the magnetization state at a given value for $D$ and the target magnetization configuration corresponding to the MFM data measured at room temperature. The evolution of the magnetization state and of the cellwise value of $|\mathbf{M} - \mathbf{M_\mathrm{target}}|$ is given in (f).}
    \label{fig:EDFig03}
\end{figure*}

\begin{figure*}
    \centering
    \includegraphics[width=0.9\textwidth]{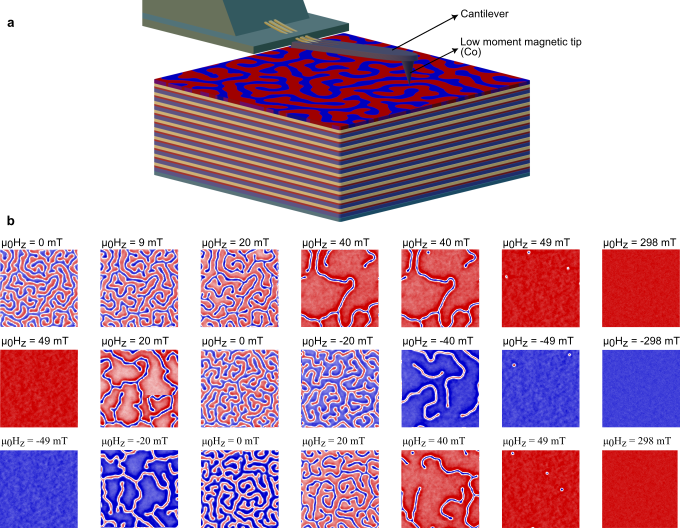}
    \caption{\textbf{Magnetic Force Microscopy Imaging (experiments)}. Schematic illustration of the measurement setup (a), where the stack is scanned with a piezoresistive cantilever that is attracted or repulsed based on the orientation of the stray-field generated by the sample. The tip of the cantilever received additional Ta(2)/Co(3)/Ta(4) deposition to enhance the sentivity and reduce the moment of the tip. The evolution of the magnetization with the applied field is given in (b), where the shrinking of stripes into skyrmions is observed for both magnetic field polarities.}
    \label{fig:EDFig04}
\end{figure*}

\begin{figure*}
    \centering
    \includegraphics[width=0.9\textwidth]{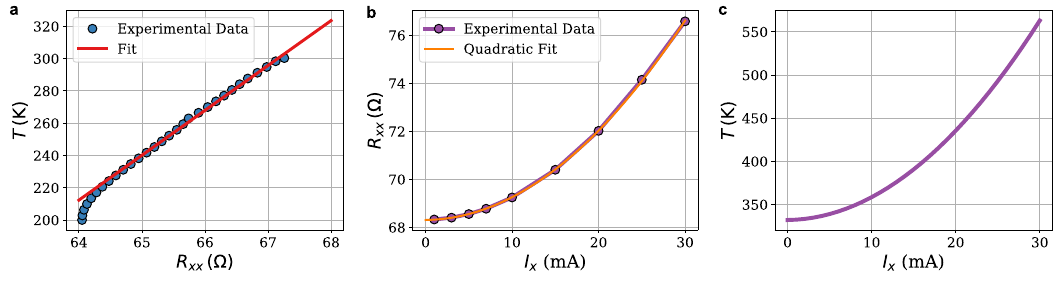}
    \caption{\textbf{Quantification of Joule heating (experiments)}. The longitudinal resistance is measured as a function of the environmental temperature set in the PPMS cryostat while the sample is saturated by an OOP field $\mu_0H_z = \SI{400}{mT}$ and a low amplitude current $I = \SI{1}{mA}$ is injected, which does not introduce any Joule heating. Then we set the environmental temperature to $T=\SI{300}{K}$, vary the applied current and record $R_{xx}$, which is depicted in (b). Ultimately, we correlate the two measurements and obtain the quadratic dependence of the sample temperature as a function of applied lateral current in (c).}
    \label{fig:EDFig05}
\end{figure*}

\begin{figure*}
    \centering
    \includegraphics[width=0.9\textwidth]{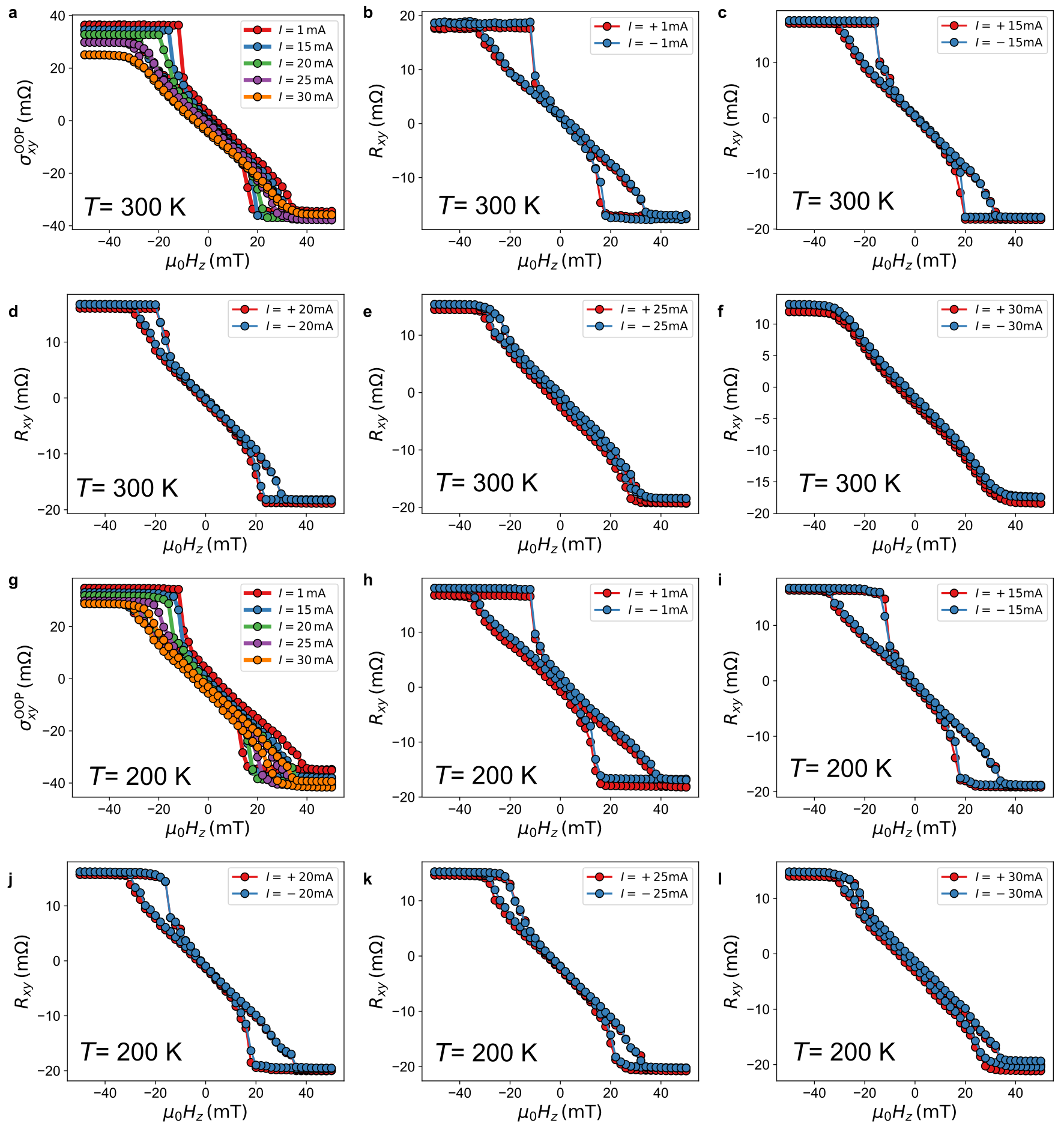}
    \caption{\textbf{Temperature and current dependence of the OOP sensing signal (experiments)}. The OOP sensing signals $\sigma^\mathrm{OOP}_{xy}$ measured for different current amplitudes in (a) at the environmental temperature $T=\SI{300}{K}$, and in (g) at $T=\SI{200}{K}$, and the individual $R_{xy}$ curves recorded for both current polarities (b) to (f) at $T=\SI{300}{K}$, and (h) to (l) at $T=\SI{200}{K}$.}
    \label{fig:EDFig06}
\end{figure*}

\begin{figure*}
    \centering
    \includegraphics[width=0.9\textwidth]{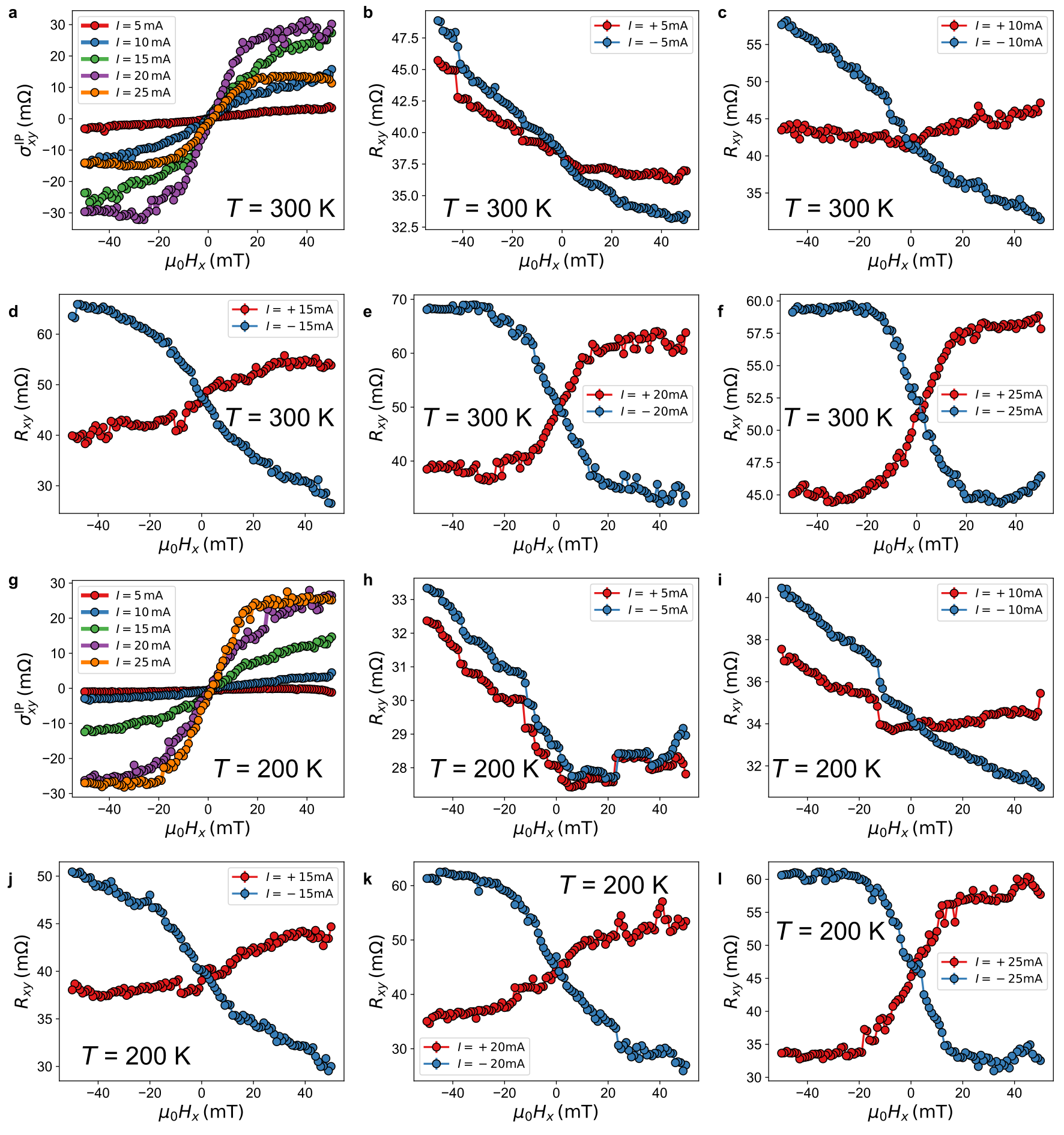}
    \caption{\textbf{Temperature and current dependence of the IP sensing signal (experiments)}. Same as Extended Data Fig.~\ref{fig:EDFig06} but for IP magnetic fields.}
    \label{fig:EDFig07}
\end{figure*}

\begin{figure*}
    \centering
    \includegraphics[width=0.9\textwidth]{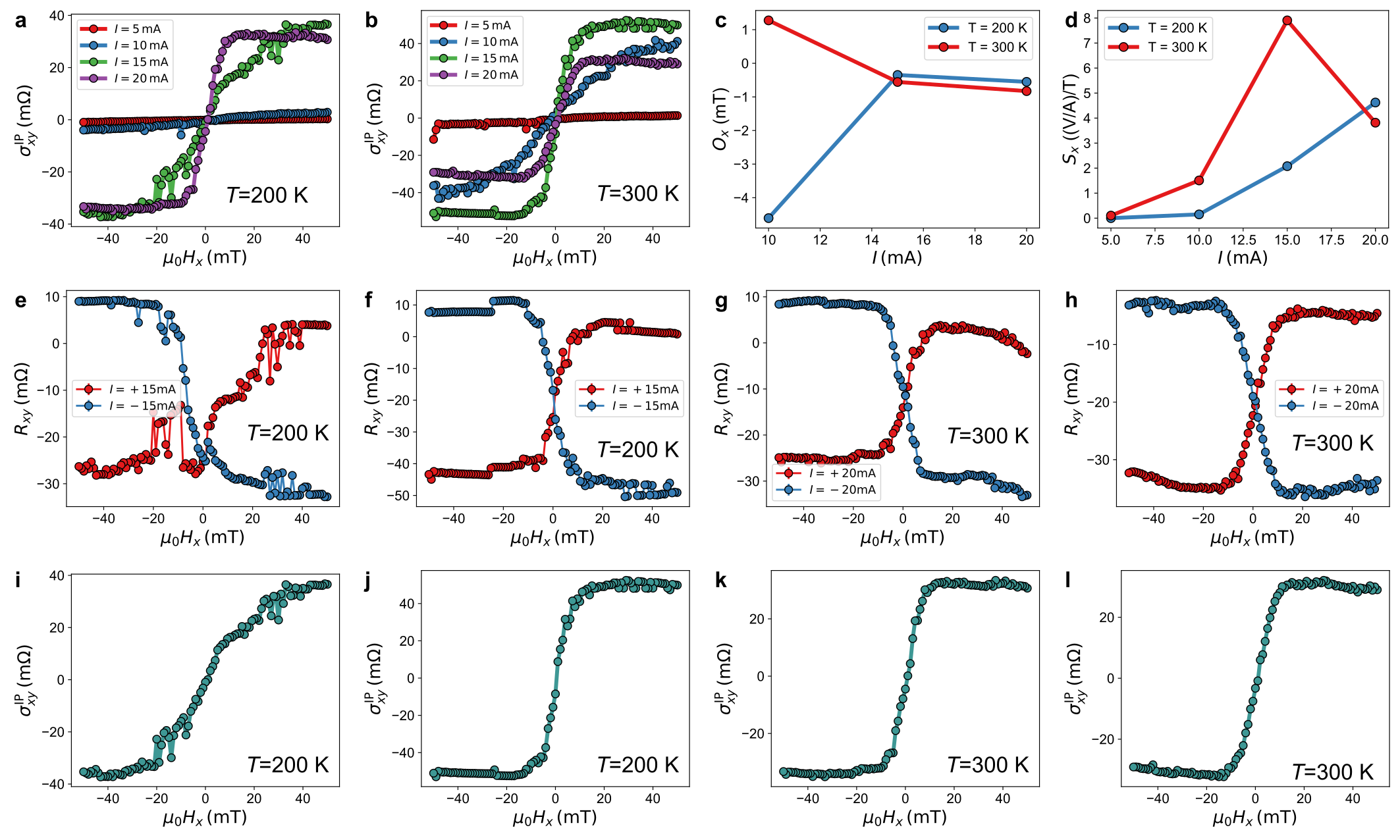}
    \caption{\textbf{Improving sensing performance in W(5)/CoFeB(0.7)/MgO(1.1)]$_\mathrm{10}$ multilayers (experiments)}. The sensor signals for an IP magnetic field are shown in (a) at $T=\SI{200}{K}$ and in (b) at $T=\SI{300}{K}$ for different lateral current amplitudes. The DC offset $O_x$ was then quantified as a function of the SOT current in (c), while the sensitivity $S_x$ is shown in (d). We observe that the sensitivity increases by a factor of 4 for the thinner MgO layer. Individual curves $R_{xy}$ are given in (e) to (f) at $T=\SI{200}{K}$, and in (g) to (h) at $T=\SI{300}{K}$, with the corresponding sensing signals in (i) to (l).}
    \label{fig:EDFig08}
\end{figure*}

\end{document}